\documentclass[twocolumn,floatfix,prc,showpacs,aps,superscriptaddress]{revtex4-1}
\usepackage{graphicx,bm}
\usepackage{amsmath}
\usepackage[colorlinks,citecolor=blue]{hyperref}
\begin{document}

\title{Astrophysical weak-interaction rates for selected $A=20$ and
  $A=24$ nuclei}

\author{G. Mart\'inez-Pinedo}
\affiliation{Institut f{\"u}r Kernphysik
  (Theoriezentrum), Technische Universit{\"a}t Darmstadt,
  Schlossgartenstra{\ss}e 2, 64289 Darmstadt, Germany}
\affiliation{GSI Helmholtzzentrum f\"ur Schwerionenforschung,
  Planckstra{\ss}e~1, 64291 Darmstadt, Germany}

\author{Y. H. Lam} 
\affiliation{Institut f{\"u}r Kernphysik
  (Theoriezentrum), Technische Universit{\"a}t Darmstadt,
  Schlossgartenstra{\ss}e 2, 64289 Darmstadt, Germany}

\author{K. Langanke}
\affiliation{GSI Helmholtzzentrum f\"ur Schwerionenforschung,
  Planckstra{\ss}e~1, 64291 Darmstadt, Germany}
\affiliation{Institut f{\"u}r Kernphysik
  (Theoriezentrum), Technische Universit{\"a}t Darmstadt,
  Schlossgartenstra{\ss}e 2, 64289 Darmstadt, Germany}
\affiliation{Frankfurt Institute for Advanced Studies, Ruth Moufang
  Str. 1, D-60438 Frankfurt, Germany}

\author{R. G. T. Zegers}
\affiliation{National Superconducting Cyclotron Laboratory, Michigan
  State University, East Lansing, Michigan 48824, USA}
\affiliation{Department of Physics and Astronomy, Michigan State
  University, East Lansing, Michigan 48824, USA}
\affiliation{Joint Institute for Nuclear Astrophysics, Michigan State
  University, East Lansing, Michigan 48824, USA} 

\author{C. Sullivan}
\affiliation{National Superconducting Cyclotron Laboratory, Michigan
  State University, East Lansing, Michigan 48824, USA}
\affiliation{Department of Physics and Astronomy, Michigan State
  University, East Lansing, Michigan 48824, USA}
\affiliation{Joint Institute for Nuclear Astrophysics, Michigan State
  University, East Lansing, Michigan 48824, USA} 

\date{\today}

\begin{abstract}
  We have evaluated the electron capture rates on $^{20}$Ne, $^{20}$F,
  $^{24}$Mg, $^{24}$Na and the $\beta$ decay rates for $^{20}$F and
  $^{24}$Na at temperature and density conditions relevant for the
  late-evolution stages of stars with $M=8$--12~M$_\odot$. The rates are
  based on recent experimental data and large-scale shell model
  calculations.  We show that the electron capture rates on $^{20}$Ne,
  $^{24}$Mg and the $^{20}$F, $^{24}$Na $\beta$-decay rates are based
  on data in this astrophysical range, except for the capture rate on
  $^{20}$Ne, which we predict to have a dominating contribution from
  the second-forbidden transition between the $^{20}$Ne and $^{20}$F
  ground states in the density range $\log \rho Y_e
  (\mathrm{g~cm}^{-3}) = 9.3$--9.6. The dominance 
  of a few individual transitions allows us to present the various rates
  by analytical expressions at the relevant astrophysical conditions.
  We also derive the screening corrections to the rates. 
\end{abstract}
\maketitle

\section{Introduction}

Electron captures on nuclei play a crucial role in the high-density
environment of late-stage stellar
evolution~\cite{Bethe:1990,Langanke.Martinez-Pinedo:2003} with three
important consequences. It reduces the pressure which the degenerate
relativistic electron gas can supply against the gravitational
contraction of the stellar core. Furthermore it cools the core
environment as the neutrinos produced in the capture process can leave
the star virtually unhindered (as long as the density is less than
about $10^{11}$~g~cm$^{-3}$ and carry away energy. Finally electron
captures change protons in the nucleus into neutrons and hence drive
the stellar composition more neutron rich.

Improving on the pioneering work by Fuller, Fowler and Newman
(FFN)~\cite{Fuller.Fowler.Newman:1980,*Fuller.Fowler.Newman:1982a,*Fuller.Fowler.Newman:1982b,Fuller.Fowler.Newman:1985}
and making use of advances in nuclear modeling and in computational
hard- and software development, electron capture rates have been
determined for sd-shell nuclei ($A=17$--39)~\cite{Oda.Hino.ea:1994}
and for pf-shell nuclei ($A=45$--64) \cite{Caurier.Langanke.ea:1999,%
  Langanke.Martinez-Pinedo:2000,Langanke.Martinez-Pinedo:2001} based
on large-scale shell-model diagonalization calculations. The
reliability of the calculations benefitted also strongly from
experimental data for the Gamow-Teller (GT$_+$) distribution in nuclei
(e.g.~\cite{frekers:2006,Fujita.Rubio.Gelletly:2011,Sasano.Perdikakis.ea:2011})
which determine the electron capture rates at the stellar conditions
for which nuclei in the mass range $A=17$--64 dominate the stellar
matter composition. Indeed a detailed comparison of stellar capture
rates derived from experimental GT$_+$ distributions for all
$pf$-shell nuclei, for which data exist, with modern shell-model rates
convincingly validated the use of the latter in late-stage stellar
evolution studies~\cite{Cole.Anderson.ea:2012} (for applications and
consequences see~\cite{Janka.Langanke.ea:2007}). We note that
diagonalization shell-model calculations are yet not globally feasible
for the very neutron-rich nuclei with $A>64$ which dominate the
electron capture at densities in excess of a few
$10^{10}$~g~cm$^{-3}$~\cite{Juodagalvis.Langanke.ea:2010} and the
respective rates must be determined based on other approaches such as
the Random Phase Approximation (RPA) with occupation numbers from
Shell Monte Carlo~\cite{Langanke.Martinez-Pinedo.ea:2003}, thermofield
dynamics approach~\cite{Dzhioev.Vdovin.ea:2010}, and finite-temperature
Quasiparticle RPA~\cite{Paar.Colo.ea:2009,Niu.Paar.ea:2011}.

While stellar electron captures usually occur on an ensemble of nuclei
present in the matter composition, capture on the specific nuclei
$^{20}$Ne and $^{24}$Mg has been identified as crucial for the core
collapse of 8--12~M$_\odot$ stars
\cite{Nomoto:1984,*Nomoto:1987,hillebrandt.nomoto.wolff:1984}. Stars in
this mass 
range develop degenerate ONe or ONeMg cores which are driven towards
collapse in a process dubbed electron capture supernova triggered by
the loss of electron pressure support due to electron captures, mainly
on the very abundant nuclear species $^{20}$Ne and $^{24}$Mg
\cite{Nomoto:1984,*Nomoto:1987,hillebrandt.nomoto.wolff:1984,Huedepohl.Mueller.ea:2010}. We
note that 8--12~M$_\odot$ stars
crucially contribute to the nucleosynthesis of specific nuclides. Its
role for the synthesis of r-process elements is currently
controversially discussed~\cite{Ning.Qian.Meyer:2007,Janka.Mueller.ea:2008}.

Simulations of late-stage evolution of 8--12~M$_\odot$ stars and
electron-capture supernovae usually adopt the weak-interaction rates,
including those for electron capture on $^{20}$Ne and $^{24}$Mg, from
the work of Oda et al. \cite{Oda.Hino.ea:1994}. These authors made
available rate tabulations for an extensive set of nuclei in the mass
range $A=17-39$, however, on a rather sparse temperature-density grid
which is argued to be insufficient for detailed studies of the
evolution stage for which the weak rates are
essential~\cite{Jones.Hirschi.ea:2013}. Shell-model rates for electron
captures on $^{20}$Ne and $^{24}$Mg had previously to the work by Oda
\emph{et al.}~\cite{Oda.Hino.ea:1994} been calculated by Takahara
\emph{et al.}~\cite{Takahara.Hino.ea:1989}. Importantly the
calculations of Takahara \emph{et al.} and of Oda \emph{et al.} had
been performed before the Gamow-Teller strength distributions for
$^{20}$Ne and $^{24}$Mg have been determined by charge-exchange
experiments. Due to the isospin symmetry of the two nuclei, this goal
could not only be achieved by techniques which determine the GT$_+$
strength distribution using $(d,{}^2\textrm{He})$ and
$(t,{}^3\mathrm{He})$ reactions, but also by those measuring the GT$_-$
distribution by $(p,n)$ and $(^3\textrm{He},t)$ reactions.  (In the latter a
neutron is changed into a proton supplying the information required
for $\beta^-$ decays). The availability of these data calls for a
reevaluation of the electron capture rates which we will present in
this manuscript. 

Besides the incorporation of recent experimental GT$_+$ data, we
improve the previous rates also in two other important aspects. At
first we point to the relevance of the ground-state-to-ground-state
transition in the capture on $^{20}$Ne which, although it is of
forbidden nature, is likely to dominate the capture rate in the
astrophysically relevant temperature-density range for the core
evolution of 8-12 M$_\odot$ stars. Secondly, we correct the capture
rates for screening effects in the dense environment (which decrease
electron capture rates, but increase the competing
$\beta$-decays). Our study is completed by a reevaluation of the rates
for electron captures and $\beta$ decays of $^{20}$F and $^{24}$Na,
which are the daughters of the electron capture processes on $^{20}$Ne
and $^{24}$Mg, respectively.
   
We note that the electron capture rate on $^{20}$Ne as well as the
$^{20}$F $\beta$-decay rate is dominated by a few transitions which are
experimentally determined, except for the forbidden
ground-state-to-ground-state transition for which only an upper limit
is known. The dominance of  a few transitions allows us to present
the rates by an analytical expression for the relevant
temperature-density region which removes uncertainties associated with
extrapolations required for rate tabulations provided on a grid.

\section{Formalism}

\subsection{Rates for electron capture, $\beta$ decay and neutrino energy loss}

We are interested in electron capture and $\beta^-$ decay rates for
temperatures $T=10^8$--$10^{10}$~K and densities
$\rho=10^8$--$10^{10}$~g~cm$^{-3}$. Under these conditions nuclei are
fully ionized and the 
electrons form a degenerate relativistic Fermi gas. Hence the rate
formalism as derived by Fuller \emph{et al.} applies
\cite{Fuller.Fowler.Newman:1980,Fuller.Fowler.Newman:1982b,Fuller.Fowler.Newman:1982a,Fuller.Fowler.Newman:1985}
which we summarize in the following.

The total rate for electron capture and $\beta^-$ decay is given by

\begin{equation}
  \label{eq:rate}
  \lambda^\alpha = \frac{1}{G(Z,A,T)} \sum_{if} (2J_i+1) \lambda^\alpha_{if}
    e^{-E_i/(kT)}  ,
\end{equation}
where the sums in $i$ and $f$ run over states in the parent and
daugther nuclei, respectively, and the superscript $\alpha$ stands for
electron capture (ec) or $\beta^-$-decay. $G(Z,A,T)=\sum_i
(2J_i+1)\exp(-E_i/(kT))$ is the partition function of the parent
nucleus. The electron capture rate from state $i$ to state $f$ is
given by:

\begin{subequations}
  \label{eq:ecrate}
  \begin{equation}
    \label{eq:ecratea}
    \lambda^{\text{ec}}_{if} = \frac{\ln 2}{K} B_{if} \Phi^{\text{ec}}(q_{if}),
  \end{equation}
  \begin{equation}
    \label{eq:ecrateb}
    \Phi^{\text{ec}}(q_{if}) = \int_{w_l}^\infty w p (q_{if}
    + w)^2 F(Z,w) S_e(w) dw;
  \end{equation}
\end{subequations}
while for $\beta^-$-decay we have:

\begin{subequations}
  \label{eq:brate}
  \begin{equation}
    \label{eq:bratea}
    \lambda^{\beta^-}_{if} = \frac{\ln 2}{K} B_{if} \Phi^{\beta}(q_{if}),
  \end{equation}
  \begin{equation}
    \label{eq:brateb}
    \Phi^{\beta}(q_{if}) = \int_1^{q_{if}} w p (q_{if}
    - w)^2 F(Z+1,w) (1-S_e(w)) dw.    
  \end{equation}
\end{subequations}

The constant $K$ can be determined from superallowed Fermi transitions
and we used $K=6144\pm 2$~s~\cite{Hardy.Towner:2009}. $w$ is the
total, rest mass plus kinetic, energy of the electron in units of $m_e
c^2$, and $p=\sqrt{w^2-1}$ is the electron momentum in units of $m_e
c$.  We have introduced the energy difference between initial and
final nuclear states, $q_{if}$, in units of $m_e c^2$

\begin{equation}
  \label{eq:qn}
  q_{if} = \frac{Q_{if}}{m_e c^2}, \quad Q_{if} = (M_pc^2 - M_dc^2 + E_i -E_f),
\end{equation}
where $M_p$, $M_d$ are the nuclear masses of the parent and daughter
nucleus, respectively, while $E_i$, $E_f$ are the excitation energies
of the initial and final states. We have calculated the nuclear masses
from the tabulated atomic masses neglecting atomic binding energies.
$w_l$ is the capture threshold total energy, rest plus kinetic, in
units of $m_e c^2$ for electron capture. Depending on the value of
$q_{if}$ one has $w_l=1$ if $q_{if} > -1$, or $w_l=|q_{if}|$ if
$q_{if} < -1$.  $S_e$ is the electron distribution function, which
for the stellar conditions, we are interested in, is given by a
Fermi-Dirac distribution with temperature $T$ and chemical potential
$\mu_e$,

\begin{equation}
  \label{eq:fermie}
  S_e (E_e) = \frac{1}{\exp\left(\frac{E_e - \mu_e}{kT}\right)+1},
\end{equation}
with $E_e = w m_e c^2$.  The chemical potential, $\mu_e$, is
determined from the density inverting the relation

\begin{equation}
  \label{eq:inverme}
  \rho Y_e = \frac{m_u}{\pi^2}\left(\frac{m_e c}{\hbar}\right)^3
  \int_0^\infty (S_e-S_p) p^2 dp,
\end{equation}
where $m_u$ is the atomic mass unit and $S_p$ is the positron
distribution which is obtained from $S_e$ by the replacement
$\mu_p=-\mu_e$. Note that the density of electron-positron pairs has
been removed in~(\ref{eq:inverme}) by forming the difference
$S_e-S_p$.

Finally, $B_{if}$ is the reduced transition probability of the nuclear
transition.  Except for the forbidden ground-state-to-ground-state
transition in $^{20}$Ne we will only consider GT contributions:

\begin{equation}
  \label{eq:bgt}
  B_{if}=B_{if}(GT) = g_A^2 \frac{\langle f||\sum_k \bm{\sigma}^k \bm{t}^k_\pm || i
    \rangle^2}{2 J_i +1}.
\end{equation}
Here the matrix element is reduced with respect to the spin operator
$\bm{\sigma}$ only (Racah convention~\cite{Edmonds:1960}) and the sum
runs over all nucleons.  For the isospin operators, $\bm{t}_\pm =
(\bm{\tau}_x \pm i \bm{\tau}_y)/2$, we use the convention $\bm{t}_+ p
= n$; thus, `$+$' refers to electron capture and `$-$' to $\beta^-$
transitions.  $g_A$ is the weak axial coupling constant, $g_A =
-1.26$. When using theoretical
Gamow-Teller matrix elements we use an effective coupling constant
$g_A^{\text{eff}} = 0.74 g_A$ to account for the observed quenching of
the GT strength in shell model 
calculations~\cite{Osterfeld:1992,Brown.Wildenthal:1988,Langanke.Dean.ea:1995,Martinez-Pinedo.Poves.ea:1996b}.

The remaining factor appearing in the phase space integrals is the
Fermi function, $F(Z,w)$, that corrects the phase space integral for
the Coulomb distortion of the electron wave function near the nucleus.

In astrophysical applications, in addition of the weak interaction
rates one is also interested in the energy loss by neutrino
emission. This can be determined by including an additional power of
the neutrino energy in equations~\eqref{eq:ecrate}
and~\eqref{eq:brate}. In this case the total neutrino energy loss rate
becomes:

\begin{equation}
  \label{eq:erate}
  \xi^\alpha = \frac{1}{G(Z,A,T)} \sum_{if} (2J_i+1) \xi^\alpha_{if}
    e^{-E_i/(kT)},
\end{equation}
and for the neutrino energy loss due to electron
capture from state $i$ to state $f$ we have:

\begin{subequations}
  \label{eq:ecerate}
  \begin{equation}
    \label{eq:eceratea}
    \xi^{\text{ec}}_{if} = \frac{(\ln 2) m_e c^2}{K} B_{if}
    \Psi^{\text{ec}}(q_{if}),
  \end{equation}
  \begin{equation}
    \label{eq:ecerateb}
    \Psi^{\text{ec}}(q_{if})  = \int_{w_l}^\infty w p (q_{if} + w)^3
    F(Z,w) S_e(w) dw;
  \end{equation}
\end{subequations}
while for $\beta^-$-decay we have:

\begin{subequations}
  \label{eq:berate}
  \begin{equation}
    \label{eq:beratea}
    \xi^{\beta^-}_{if} = \frac{(\ln 2) m_e c^2}{K} B_{if} \Psi^{\beta}(q_{if}),
  \end{equation}
  \begin{equation}
    \label{eq:berateb}
    \Psi^{\beta}(q_{if}) = \int_1^{q_{if}} w p (q_{if} - w)^3 F(Z+1,w)
    (1-S_e(w)) dw. 
  \end{equation}
\end{subequations}  
Similarly one can compute the average energy of the emitted neutrino
by the ratio:

\begin{equation}
  \label{eq:nuave}
  \langle E_\nu\rangle^{\alpha} = \frac{\xi^\alpha}{\lambda^\alpha}
\end{equation}
where $\alpha$ stands for either electron capture or beta-decay.

\subsection{Approximate expressions}
\label{sec:appr-expr}

The evaluation of electron capture and beta-decay rates requires the
calculation of the phase space integrals appearing in
equations~(\ref{eq:ecrate}), (\ref{eq:brate}), (\ref{eq:ecerate}) and
(\ref{eq:berate}). These integrals make the rates extremely sensitive
to variations of temperature and density. As weak
interaction rate tabulations~\cite{Fuller.Fowler.Newman:1982a,%
Langanke.Martinez-Pinedo:2001,Oda.Hino.ea:1994} are normally provided
on a grid of densities and temperatures, this requires the development
of accurate interpolation schemes between the grid points at which the
rates have been evaluated. For the high temperature ($T>10^9$~K) and density
conditions relevant for presupernova
evolution~\cite{Heger.Langanke.ea:2001}, many transitions from both
the initial and final nucleus contribute to the sum in equation
(\ref{eq:rate}) and the density and temperature dependence of the
rates can well be aproximated by an effective phase space integral,
$\Phi^{\text{ec}}_e$, corresponding to the ground-state to
ground-state transition. This allows to introduce an effective $\langle
ft\rangle$ value that is expected to be almost constant over a large range of
temperature and densities~\cite{Fuller.Fowler.Newman:1985}:

\begin{equation}
  \label{eq:efflogft}
  \lambda^{ec} = \ln 2 \frac{\Phi^{\text{ec}}_e}{\langle ft\rangle}.  
\end{equation}

In the present work, we are interested in conditions for which URCA
processes operate in both intermediate mass
stars~\cite{Tsuruta.Cameron:1970} and neutron star
crust~\cite{Schatz.Gupta.ea:2014}. This corresponds to temperatures in
the range $10^8$--$10^9$~K for which, in contrast to the presupernova
conditions, both electron capture and beta-decay rates are determined
by a few transitions (see section~\ref{sec:results}) with very
different phase space dependencies. The application of the above
approach will result in changes in the effective $\langle ft\rangle$
value of several orders of magnitude when the rate changes from being
dominated by one transition to another transition. As this occurs in a
very narrow density range, it makes it impractical to use the effective
$\langle ft\rangle$ formalism for the rate interpolation.
Nevertheless, one can still use the fact that the rates are determined
by a few transitions to provide accurate analytic expressions for the
relevant rates. This constitutes a generalization of the FFN effective
$\langle ft\rangle$ formalism and its extension to beta-decay rates.

In the evaluation of the phase space integral in
equation~\eqref{eq:efflogft}  one can use the fact 
that for the large electron energies involved the Fermi function,
$F(Z,w)$, can be approximated up to a constant factor (which can be
subsumed in a redefinition of the matrix element) by the ratio $w/p$: 

\begin{equation}
  \label{eq:effphaseec}
  \Phi^{\text{ec}}_e(Q,T,\mu_e) = \int_{w_l}^\infty w^2 (q + w)^2 S_e(w) dw,
\end{equation}
where we have made explicit the dependence of the integral on
temperature, electron chemical potential, $\mu_e$, and $Q$-value for the
ground-state to ground-state transition, $Q = q m_e c^2$. $\Phi^{\text{ec}}_e$ can be
expressed as a combination of relativistic Fermi integrals (or
equivalently polylogarithmic functions, $\text{Li}$):
\begin{equation}
  \label{eq:fermipoly}
  \begin{aligned}
    F_k(\eta) &= \int_0^\infty \frac{x^k}{\exp(x-\eta)+1} dx,\\[2mm]
    F_k(\eta) & = - \Gamma(k+1) \text{Li}_{k+1} (-e^\eta);  
  \end{aligned}
\end{equation}
to obtain
\begin{equation}
  \label{eq:ephasefermi}
  \Phi^{\text{ec}}_e(Q,T,\mu_e) = \left(\frac{kT}{m_e c^2}\right)^5
    \left[F_4(\eta)-2 \chi F_3(\eta)+\chi^2 F_2(\eta)\right]  
\end{equation}
with $\eta = (\mu_e + Q)/(k T)$, $\chi = Q/(kT)$, and we have assumed
that $Q<-m_e c^2$, which is the case for all the nuclei considered
here. For the evaluation of the Fermi functions appearing in
equation~\eqref{eq:ephasefermi} one can use several publicly available
numerical routines both in C~\cite{Galassi.Davies.ea:2009} and in
Fortran~\cite{Aparicio:1998,Gong.Zejda.ea:2001}. Alternatively,
\citeauthor{Fuller.Fowler.Newman:1985}~\cite{Fuller.Fowler.Newman:1985}
developed approximations for the Fermi functions that are valid for
$\eta \ll 0$ and $\eta \gg 0$ and reproduce the exact results with an
accuracy of better than 20\% around $\eta \approx 0$:

\begin{subequations}
  \label{eq:fermiapprox}
  \begin{equation}
    \label{eq:fermi0}
    F_0(\eta) = \ln \left(1+e^\eta\right);
  \end{equation}
  \begin{equation}
    \label{eq:fermi1}
    F_1(\eta)=\begin{cases}
      e^{\eta } & \eta \leq 0, \\
      \frac{1}{2}\eta^2+2-e^{-\eta } & \eta >0;
    \end{cases}
  \end{equation}
  \begin{equation}
    \label{eq:fermi2}
    F_2(\eta)=\begin{cases}
      2 e^{\eta } & \eta \leq 0, \\
      \frac{1}{3}\eta^3+4 \eta +2 e^{-\eta } & \eta >0;
    \end{cases}
  \end{equation}
  \begin{equation}
    \label{eq:fermi3}
    F_3(\eta)=\begin{cases}
      6 e^{\eta } & \eta \leq 0, \\
      \frac{1}{4}\eta^4+\frac{\pi^2}{2}\eta^2+12-6 e^{-\eta } & \eta >0;
    \end{cases}
  \end{equation}
  \begin{equation}
    \label{eq:fermi4}
    F_4(\eta)=\begin{cases}
      24 e^{\eta } & \eta \leq 0, \\
      \frac{1}{5}\eta^5+\frac{2 \pi^2}{3} \eta^3+48 \eta +24
      e^{-\eta } & \eta >0; 
    \end{cases}
  \end{equation}
  \begin{equation}
    \label{eq:fermi5}
    F_5(\eta)=\begin{cases}
      120 e^{\eta} & \eta \leq 0, \\
      \frac{1}{6}\eta^6+\frac{5 \pi^2}{6} \eta^4+\frac{7
        \pi^4}{6}\eta^2+240-120 e^{-\eta }  & \eta >0;
    \end{cases}
  \end{equation}
\end{subequations}
where we have included the expressions for $F_0(\eta)$ and $F_1(\eta)$
that are necessary for the beta-decay rates and corrected for a typo
in the approximation of $F_5(\eta)$ in
ref.~\cite{Fuller.Fowler.Newman:1985}.  

The partial contribution to the total electron capture rate of an initial
state, $i$, with excitation energy $E_i$ and angular momentum $J_i$ to
a final state, $f$, with excitation energy $E_f$ and angular momentum
$J_f$ can be expressed as:

\begin{widetext}
\begin{equation}
  \label{eq:contij}
  \Lambda^{\text{ec}}_{if} = (2J_i+1) e^{-E_i/(k T)} \lambda_{if}^{\text{ec}} = \frac{(\ln 2) B^e_{if}}{K} (2 J_i+1) e^{-E_i/(k T)}
  \Phi_e^{\text{ec}}(Q_{if},T,\mu_e)   
\end{equation}
\end{widetext}
with $Q_{if} = Q+E_i-E_f$. During the early evolution of an ONeMg core
the electron chemical potential, $\mu_e$, is typically much smaller
than the magnitude of the capture $Q$-value, $|Q_{if}|$, i.e. $\eta \ll
0$. Under these conditions the Fermi integrals appearing
in~\eqref{eq:ephasefermi} can be approximated as $F_k(\eta) \approx
k! e^\eta$. Keeping the leading terms in eq.~\eqref{eq:contij} we obtain:

\begin{widetext}
  \begin{equation}
  \label{eq:contijse}
  \Lambda^{\text{ec}}_{if} = \frac{(\ln 2) B^e_{if}}{K}
  \left(\frac{kT}{m_e c^2}\right)^5 2 (2 J_i+1)
  \left(\frac{Q+E_i-E_f}{k T}\right)^2 \exp\left(\frac{Q-E_f+\mu_e}{k
      T}\right)    
\end{equation}
\end{widetext}

One can see that the exponential dependence on the excitation energy
of the initial state has disappeared. The physical reason is that with
increasing excitation energy the exponential decrease in the thermal
probability of populating an excited state is exactly compensated by
the exponential increase in the number of electrons that can
contribute to the capture process. Under these conditions the rate
grows exponentially with increasing electron chemical potential. This
increase holds as long as $\mu_e \ll -Q_{if}$ and $E_i \ll -Q$. Once
the chemical potential $\mu_e$ becomes larger than the absolute 
$Q$-value, the Fermi integrals can be approximated as $F_k(\eta) \approx
\eta^{k+1}/(k+1)$. It is interesting to consider two possible limits,
i) the electron fermi energy is similar to the capture $Q$-value,
$\mu_e \approx |Q_{if}|$ and ii) the electron fermi energy is much
larger than the capture $Q$-value, $\mu_e \gg |Q_{if}|$. In the first
case we obtain

\begin{widetext}
  \begin{equation}
    \label{eq::contijme}
    \Lambda^{\text{ec}}_{if} = \frac{(\ln 2) B^e_{if}}{3 K}(2
    J_i+1)\exp\left(-\frac{E_i}{k T}\right) \frac{(Q+E_i-E_f)^2
    (\mu_e +Q +E_i -E_f)^3}{(m_e c^2)^5},
  \end{equation}
while for the second  
  \begin{equation}
    \label{eq:contijle}
    \Lambda^{\text{ec}}_{if} = \frac{(\ln 2) B^e_{if}}{5 K} (2
    J_i+1)\exp\left(-\frac{E_i}{k T}\right) \left(\frac{\mu_e +Q +E_i
        -E_f}{m_e c^2}\right)^5. 
  \end{equation}
\end{widetext}
Under these conditions the contribution of excited states is
exponentially suppressed and the capture rate on each state is almost
independent of the temperature. 

Similar approximations can be obtained for the neutrino energy loss
rate. The contribution of a transition from an initial state $i$ to a
final state $f$ is then given by:

\begin{widetext}
\begin{equation}
  \label{eq:contelossij}
  \Xi^{\text{ec}}_{if} = (2J_i+1) \xi^\alpha_{if}
    e^{-E_i/(kT)} = \frac{(\ln 2) B^e_{if}}{K} m_e c^2 (2J_i+1)e^{-E_i/(kT)}
  \Psi^{\text{ec}}_e(Q_{if},T,\mu_e)  
\end{equation}
with
\begin{equation}
  \label{eq:lossfermi}
  \Psi^{\text{ec}}_e(Q,T,\mu_e) = \left(\frac{k T}{m_e c^2}\right)^6
       \left[F_5(\eta)-2\chi F_4(\eta) + \chi^2 F_3(\eta)\right],
\end{equation}
and $\eta = (\mu_e + Q)/(k T)$, $\chi = Q/(kT)$. 

Again we can obtain approximate expressions for the  limiting cases
$\mu_e \ll -Q_{if}$:

\begin{equation}
  \label{eq:elossse}
  \Xi^{\text{ec}}_{if} = \frac{(\ln 2) B^e_{if}}{K} 
  \frac{(kT)^6}{(m_e c^2)^5} 6 (2
  J_i+1)\left(\frac{Q+E_i-E_f}{k T}\right)^2 
  \exp\left(\frac{Q-E_f+\mu_e}{k T}\right);
\end{equation}
$\mu_e \approx -Q_{if}$:
\begin{equation}
  \label{eq:elossme}
  \Xi^{\text{ec}}_{if}=\frac{(\ln 2) B^e_{if}}{4 K} (2
  J_i+1)\exp\left(-\frac{E_i}{k T}\right) \frac{(Q+E_i-E_f)^2(\mu_e +Q +E_i
      -E_f)^4}{(m_e c^2)^5};
\end{equation}
and $\mu_e \gg -Q_{if}$:
\begin{equation}
  \label{eq:elossle}
  \Xi^{\text{ec}}_{if}=\frac{(\ln 2) B^e_{if}}{6 K} (2
  J_i+1)\exp\left(-\frac{E_i}{k T}\right) \frac{(\mu_e +Q +E_i
      -E_f)^6}{(m_e c^2)^5}.
\end{equation}
\end{widetext}
Combining equations~\eqref{eq:contijse} and~\eqref{eq:elossse} one
obtains that the average neutrino energy for conditions
$\mu_e \ll -Q_{if}$ is:
\begin{equation}
  \label{eq:enulossse}
  \langle E_\nu\rangle^{\text{ec}} \approx 3 k T
\end{equation}
independently of the initial state on which the electron capture takes
place. Similarly from equations~\eqref{eq::contijme} and
\eqref{eq:elossme} we obtain for $\mu_e \approx -Q_{if}$:

\begin{equation}
  \label{eq:enulossme}
   \langle E_\nu\rangle^{\text{ec}} = \frac{3}{4} (\mu_e +Q +E_i
        -E_f)  
\end{equation}
that agrees with the result of refs.~\cite{Haensel.Zdunik:2003}
and~\cite{Bethe.Brown.ea:1979} (but with a factor 3/4 instead of
3/5). From equations~\eqref{eq:contijle} and \eqref{eq:elossle} we
obtain for $\mu_e \gg -Q_{if}$ 

\begin{equation}
  \label{eq:enulossle}
   \langle E_\nu\rangle^{\text{ec}} = \frac{5}{6} (\mu_e +Q +E_i
        -E_f) \approx \frac{5}{6} \mu_e
\end{equation}
recovering the well known result of
refs.~\cite{Bethe.Brown.ea:1979,Shapiro.Teukolsky:1983}.

The beta-decay rates can also be expressed as combinations of Fermi
functions. For that one can approximate the phase space integral in
equation~\eqref{eq:brateb} by:

\begin{equation}
  \label{eq:betaphase}
  \Phi_e^{\beta}(Q,T,\mu_e) = \int_1^{q} w^2 (q-w)^2 (1-S_e(w)) dw 
\end{equation}
with $q=Q/(m_e c^2)$. This can be expressed in terms of Fermi
functions as:

\begin{widetext}
  \begin{equation}
    \label{eq:betaphasefermi}
    \begin{split}
      \Phi_e^{\beta}(Q,T,\mu_e) = \left(\frac{k T}{m_e c^2}\right)^5
      \bigl[&-\vartheta ^2 (\vartheta -\chi )^2
      F_0\left(-\eta_m\right) + 2 \vartheta \left(\chi ^2+2 \vartheta
        ^2-3 \chi\vartheta \right)
      F_1\left(-\eta_m\right)+\left(-\chi^2-6 \vartheta ^2+6 \chi
        \vartheta \right) F_2\left(-\eta
        _m\right)\\
      & +(4 \vartheta -2 \chi ) F_3\left(-\eta
        _m\right)-F_4\left(-\eta _m\right)+\chi ^2
      F_2\left(-\eta\right)-2 \chi
      F_3\left(-\eta\right)+F_4\left(-\eta\right)\bigr]
    \end{split}
  \end{equation}
\end{widetext}

\noindent
with $\vartheta = m_e c^2/(k T)$, $\chi = Q/(k T)$, $\eta_m = (\mu_e -
m_e c^2)/(k T)$ and $\eta = (\mu_e-Q)/(k T)$. For the conditions we
are interested in one has $\eta_m \gg 1$. The Fermi integrals with arguments 
$-\eta_m$ behave like $\exp(-\eta_m)$ and their contributions can be
neglected. Under these conditions we obtain:

\begin{equation}
  \label{eq:betaphasefermiapp}
  \Phi_e^{\beta}(Q,T,\mu_e) = \left(\frac{k T}{m_e c^2}\right)^5
  \left[F_4\left(-\eta\right)-2 \chi
    F_3\left(-\eta\right)+\chi^2 F_2\left(-\eta\right)\right],
\end{equation}
which is very similar to equation~\eqref{eq:ephasefermi}.  In the
limit where final state blocking can be neglected, i.e. $\mu_e \ll m_e c^2$,
equation~\eqref{eq:betaphasefermi} reduces to:
\begin{equation}
  \label{eq:1}
  \Phi_e^{\beta}(Q,T,\mu_e) \approx \frac{1}{30} (q-1)^3 (6 + 3
    q+q^2) 
\end{equation}
with $q=Q/(m_e c^2)$.
For partial blocking of the final state, $\mu_e \lesssim Q$, we obtain

\begin{equation}
  \label{eq:2}
  \Phi_e^{\beta}(Q,T,\mu_e) \approx \frac{Q^2 (Q-\mu_e)^3}{3(m_e
    c^2)^5}   
\end{equation}

In the limit of strong final state blocking,
i.e. $\mu_e \gg Q$ we get:
\begin{equation}
  \label{eq:3}
  \Phi_e^{\beta}(Q,T,\mu_e) \approx 2\left(\frac{k T}{m_e
      c^2}\right)^5
  \left(\frac{Q}{kT}\right)^2\exp\left(\frac{Q-\mu_e}{kT}\right).
\end{equation} 
For these conditions, the contribution of a transition from an initial state $i$
to a final state $f$ to the beta-decay rate can be expressed as:

\begin{widetext}
  \begin{equation}
    \label{eq:contbetaif}
    \Lambda_{if}^{\beta}=\frac{(\ln 2) B^e_{if}}{K} \left(\frac{k T}{m_e
      c^2}\right)^5 2  (2J_i+1) \left(\frac{Q+E_i-E_f}{k T}\right)^2
    \exp\left(\frac{Q-E_f -\mu_e}{kT}\right)
  \end{equation}
\end{widetext}
As for electron capture, the $\beta$-decay rate
does not
depend on the excitation energy of the initial state. Furthermore, the strong similarity with
equation~\eqref{eq:contijse} is remarkable. This shows that beta-decays decrease
with exactly the same exponential dependence on $\mu_e$ as the
electron captures increase and probes the potential of equation
\eqref{eq:betaphasefermi} for interpolating beta-decay rates under
presupernova conditions. 

For the rate of neutrino energy loss by beta-decay we obtain:

\begin{widetext}
  \begin{equation}
    \label{eq:betaephasefermi}
    \begin{split}
      \Psi_e^{\beta}(Q,T,\mu_e) =  \left(\frac{k T}{m_e c^2}\right)^6
    \bigl[&\vartheta
   ^2 (\vartheta -\chi )^3 F_0\left(-\eta_m\right)-\vartheta  (5
   \vartheta -2 \chi ) (\vartheta -\chi )^2 F_1\left(-\eta
   _m\right)\\
 &+\left(\chi^2+10
     \vartheta ^2-8 \chi\vartheta \right) (\vartheta  -\chi )
   F_2\left(-\eta _m\right)
   -\left(3 \chi ^2+10 \vartheta^2-12 \chi\vartheta\right)
   F_3\left(-\eta _m\right)\\
   &+(5 \vartheta -3 \chi )
   F_4\left(-\eta_m\right)-F_5\left(-\eta_m\right)
   +\chi^2
   F_3\left(-\eta\right)
   -2 \chi  F_4\left(-\eta\right)+F_5\left(-\eta\right)\bigr].
 \end{split}
\end{equation}
\end{widetext}
For $\eta_m \gg 1$ the expression reads

\begin{equation}
  \label{eq:betaephasefermiapp}
  \Psi_e^{\beta}(Q,T,\mu_e) =  \left(\frac{k T}{m_e c^2}\right)^6
  \left[F_5\left(-\eta\right) -2 \chi  F_4\left(-\eta
      \right)+ \chi ^2 F_3\left(-\eta\right)\right]
\end{equation}

In the limit of no final state blocking, i.e. $\mu_e \ll m_e c^2$,
eq.~\eqref{eq:betaephasefermi} reduces to:
\begin{equation}
  \label{eq:4}
  \Psi_e^\beta(Q, T,\mu_e) \approx \frac{1}{60} (q-1)^4 (q^2+4 q + 10).
\end{equation}
The average energy of the emitted neutrino becomes:
\begin{equation}
  \label{eq:5}
  \langle E_\nu \rangle^\beta = m_e c^2 \frac{(q-1)(q^2+4 q+ 10)}{2 (q^2 + 3
    q + 6)} \approx m_e c^2 \left(\frac{q}{2} - \frac{5}{q^2}\right).
\end{equation}

For partial final state blocking, $\mu_e \lesssim Q$ we get:
\begin{equation}
  \label{eq:6}
  \Psi_e^{\beta}(Q,T,\mu_e) \approx \frac{Q^2 (Q-\mu_e)^4}{4 m_e c^2} 
\end{equation}
and for the average energy of the emitted neutrino:
\begin{equation}
  \label{eq:enubetamultq}
  \langle E_\nu \rangle^\beta = \frac{3}{4} (Q+E_i -E_f -\mu_e),
\end{equation}
where we have explicitly recovered the dependence on the excitation
energies of initial and final states to make clear the similarity with
equation~\eqref{eq:enulossme} for electron capture. 

In the limit of large final state blocking, i.e. $\mu_e \gg Q$ we
obtain:
\begin{equation}
  \label{eq:8}
    \Psi_e^{\beta}(Q,T,\mu_e) \approx 6\left(\frac{k T}{m_e
      c^2}\right)^6  \left(\frac{Q}{kT}\right)^2
  \exp\left(\frac{Q-\mu_e}{kT}\right) 
\end{equation}
and for the contribution to the total beta-decay rate of the
transition $i \rightarrow f$:

\begin{widetext}
\begin{equation}
  \label{eq:betalossse}
  \Xi_{if}^\beta =\frac{(\ln 2) B^e_{if}}{K} m_e c^2 \left(\frac{kT}{m_e c^2}\right)^6
  6 (2 J_i+1)\left(\frac{Q+E_i-E_f}{k T}\right)^2
  \exp\left(\frac{Q-E_f-\mu_e}{k T}\right)  .
\end{equation}
\end{widetext}
This again is remarkably similar to the equivalent expression for
electron capture~\eqref{eq:elossse}. Under these conditions the average
energy of the emitted neutrino becomes:

\begin{equation}
  \label{eq:9}
  \langle E_\nu \rangle^\beta = 3 k T
\end{equation}
and is independent of the particular transitions that dominate the
rate. 

\subsection{Determination of energy generation}
\label{sec:determ-energy-gener}

Apart of the fact that weak interaction processes change the electron
content of the star, they are also important because they can be
either a source or loss of energy for the star. The neutrinos that are
produced by the weak interaction leave the star carrying away part of
the energy generated. However, depending on the conditions the net
energy generation could still be positive or negative. From basic
thermodynamics and assuming that the time scale to maintain
thermodynamical equilibrium is shorter than the time scale for weak
interaction processes, we obtain the following relation valid at every
point of the star:

\begin{equation}
  \label{eq:entropy}
  k T \frac{ds}{dt} + \sum_i \mu_i \frac{dY_i}{dt} = \frac{dq}{dt}
\end{equation}
where $s$ is the entropy per nucleon, $\mu_i$ is the chemical
potential including rest mass for species $i$ with abundance $Y_i$ and
the sum runs over all particles including nuclei and
electrons. $dq/dt$ represents the heat per nucleon and time which is being added to or
lost from the region being considered. In the case of weak processes,
heat is lost by neutrinos. In the case of electron capture in a
nucleus $a$ producing a nucleus $b$ with energy threshold
$Q_{\text{ec}} = -Q_{\beta^-} = -Q$, i.e. the ground-state to
ground-state $Q$-value, we have:

\begin{equation}
  \label{eq:entropyec}
  k T \frac{ds}{dt} = -\frac{dY_e}{dt} (\mu_e - Q -
  \langle E_\nu \rangle^{\text{ec}})  - kT \frac{dY_e}{dt} \ln\left[\frac{Y_a
      G_b}{Y_b G_a }\right]
\end{equation}
while for the beta decay of nucleus $b$ to a nucleus $a$ we obtain
\begin{equation}
  \label{eq:entropybeta}
  k T \frac{ds}{dt} = \frac{dY_e}{dt} (Q - \mu_e -
  \langle E_\nu \rangle^{\beta})  + kT \frac{dY_e}{dt} \ln\left[\frac{Y_b
      G_a}{Y_a G_b }\right],  
\end{equation}
where $Y_{a,b}$ represent the abundances of nuclei $a,b$ and the
second term in both expressions has been obtained assuming a
non-interacting Boltzmann gas expression for the chemical potential of
nuclei. This term is typically negligible except in the case of very
different abundance of nuclei $a$ and $b$.

Equations~\eqref{eq:entropyec} and~\eqref{eq:entropybeta} show that the
determination of the energy generation requires only the knowledge of
the average energy of the produced neutrino and it is not necessary to
compute the so-called gamma ray heating
rates~\cite{Takahara.Hino.ea:1989,Oda.Hino.ea:1994}, corresponding to
the decay by gamma emission of excited states populated after the weak
transition. This does not mean that transitions to excited states are
not important. In fact, they normally have larger contributions to the
energy generation~\cite{Gupta.Brown.ea:2007} due to the fact that the
average energy of the neutrinos is smaller. However, one has also to
consider that the weak process can start in excited states and in that
case the gamma heating rates can in fact be negative. 

As electron captures decrease $Y_e$, while beta decays increase $Y_e$,
the energy generation will be positive or negative depending on the
sign of the quantity $\mathcal{E}^{\text{ec}} = \mu_e - Q - \langle
E_\nu \rangle^{\text{ec}}$ for electron capture and
$\mathcal{E}^{\beta} = Q -\mu_e - \langle E_\nu \rangle^{\beta}$ for
$\beta^-$-decay (neglecting the second terms in
Eqs.~\eqref{eq:entropyec} and ~\eqref{eq:entropybeta}). Using energy
conservation to relate the energy of the electron and the energy of
the neutrino in equations~\eqref{eq:ecerateb} and \eqref{eq:berateb}
we obtain the following relations for electron capture,

\begin{equation}
  \label{eq:enerconsec}
  \langle E_e\rangle^{\text{ec}} + \langle E_i\rangle^{\text{ec}} =
  \langle E_f\rangle^{\text{ec}}  + Q + \langle E_\nu \rangle^{\text{ec}}, 
\end{equation}
and beta decay,
\begin{equation}
  \label{eq:betaconsec}
  \langle E_i\rangle^{\beta} + Q  = \langle E_f\rangle^{\beta}  + \langle
  E_e\rangle^{\beta} + \langle E_\nu \rangle^{\beta},
\end{equation}
where $\langle E_e\rangle^{\text{ec}}$ ($\langle E_e\rangle^{\beta}$)
is the average energy of the captured electron (emitted electron) and
$\langle E_i\rangle^{\text{ec}}$ and $\langle E_f\rangle^{\text{ec}}$
($\langle E_i\rangle^{\beta}$ and $\langle E_f^{\beta}\rangle$)
represent the average energy of the initial and final nuclei in the
electron capture ($\beta$-decay) process.  We can define the average
energy of the produced gammas as: $\langle E_\gamma \rangle^\alpha =
\langle E_f\rangle^\alpha - \langle E_i\rangle^\alpha$, where $\alpha$
stands for electron capture and beta decay. Notice that this quantity
can be negative, meaning that transitions from excited parent nuclear
states dominate the weak process and the nucleus has to absorb gamma
radiation to populate these states. Combining the above expressions we
obtain the following relations for the quantities
$\mathcal{E}^{\text{ec}}$ and $\mathcal{E}^{\beta}$:

\begin{subequations}
  \label{eq:egeneration}
  \begin{equation}
    \label{eq:egeneec}
    \mathcal{E}^{\text{ec}} = \mu_e - Q - \langle E_\nu
\rangle^{\text{ec}} =  \mu_e - \langle E_e\rangle^{\text{ec}} +
\langle E_\gamma\rangle^{\text{ec}} 
  \end{equation}
  \begin{equation}
    \label{eq:egenebeta}
    \mathcal{E}^{\beta} = Q - \mu_e - \langle E_\nu
\rangle^{\beta} =  \langle E_e\rangle^{\beta} + 
\langle E_\gamma\rangle^{\beta} - \mu_e 
  \end{equation}
\end{subequations}

Electron captures will be endothermic, i.e. they absorb energy,
when the electron chemical potential is smaller than
$Q$. Under these conditions $\langle E_\nu \rangle^{\text{ec}} \approx
3 kT$, see eq~\eqref{eq:enulossse}, and $\mathcal{E}^{\text{ec}}
\approx -(Q-\mu_e+3kT)$. Equivalently, this means that the average
energy of the captured electrons is larger than the sum of the
chemical potential and gamma ray
energies~\cite{Miyaji.Nomoto.ea:1980}. Under these conditions the
electron capture rate is rather small and beta decay is exothermic,
i.e. it generates energy, with 

\begin{equation}
  \label{eq:7}
  \mathcal{E}^{\beta} \approx \langle
  E_\gamma\rangle^{\beta} + (Q-\langle E_\gamma\rangle^{\beta}-\mu_e)/4
\end{equation}
where we have used \eqref{eq:enubetamultq} to relate the neutrino
energy loss rate to the average gamma energy. Equivalently, see
eq.~\eqref{eq:egenebeta}, the sum of the average energies of the
electrons and gammas produced by beta decay is larger than the
electron chemical potential. Under these conditions, the beta decay
rate is normally much larger than the electron capture rate, however,
the abundance of the beta-decaying nucleus is still rather small as
very little material has been produced by electron capture.

As the electron chemical potential grows, the net energy generation in
electron capture and beta-decay could be positive or negative
depending on the particular temperature/density conditions and the
structure of the nuclei involved.

Electron captures will be exothermic whenever the electron chemical
potential becomes larger than the capture threshold. For these
conditions the electron capture proceeds rapidly and the heating is
large with 

\begin{equation}
  \label{eq:energenec}
  \mathcal{E}^{\text{ec}} \approx \langle
  E_\gamma\rangle^{\text{ec}} + (\mu_e - Q - \langle
  E_\gamma\rangle^{\text{ec}})/4,
\end{equation}
where we have estimated the neutrino energy loss using
eq.~\eqref{eq:enulossme}. The energy generation becomes larger the
higher the average energy of the produced
gammas~\cite{Gupta.Brown.ea:2007}, i.e. the higher the excitation
energy of the final states. The electrons captured have on average
energies smaller than the sum of the electron chemical potential plus
the average gamma energy~\cite{Nakazawa:1973}, see
equation~\eqref{eq:egeneec}. Under these conditions the beta-decay
rate is rather small and its contribution to the energy generation is
negligible.

\section{Results}
\label{sec:results}

\subsection{Rates for the $A=20$ nuclei}

For the study of the electron capture on $^{20}$Ne and the $\beta^-$
decay of $^{20}$F we have adopted the following set of experimental
and calculated transitions. We use the $(p,n)$ data on $^{20}$Ne
of~\cite{Anderson.Tamimi.ea:1991} and assume isospin symmetry to
determine the GT$_+$ transitions from the $^{20}$Ne ground state to
low lying excited $1^+$ states in $^{20}$F. From $^{20}$F $\beta$
decay data~\cite{Ajzenberg-Selove:1987} we determined the GT matrix
element for the transition from the $2^+$ state in $^{20}$Ne at
$E_x=1.634$~MeV to the $2^+$ $^{20}$F ground state. We approximate the
non-unique second forbidden transition from the $2^+$ ground state in
$^{20}$Ne to the $2^+$ ground state in $^{20}$F by the upper limit
obtained in the $^{20}$F $\beta$ data. 

The experimental data were supplemented by shell model GT$_+$ strength
functions from the $^{20}$Ne first $2^+$ and $4^+$ excited states, and
the backresonance transitions corresponding to the $GT_-$ strength on
the $2^+$ ground state and the lowest $3^+$, $4^+$, $1^+$, $5^+$ and $2^+$
excited states of $^{20}$F. The shell-model calculations were
performed within the complete $sd$-shell using the USDB
interaction~\cite{Brown.Richter:2006}. 
We use experimental values for the excitation energies whenever they
are known.
 
\begin{figure}[htb]
  \centering
  \includegraphics[width=\linewidth]{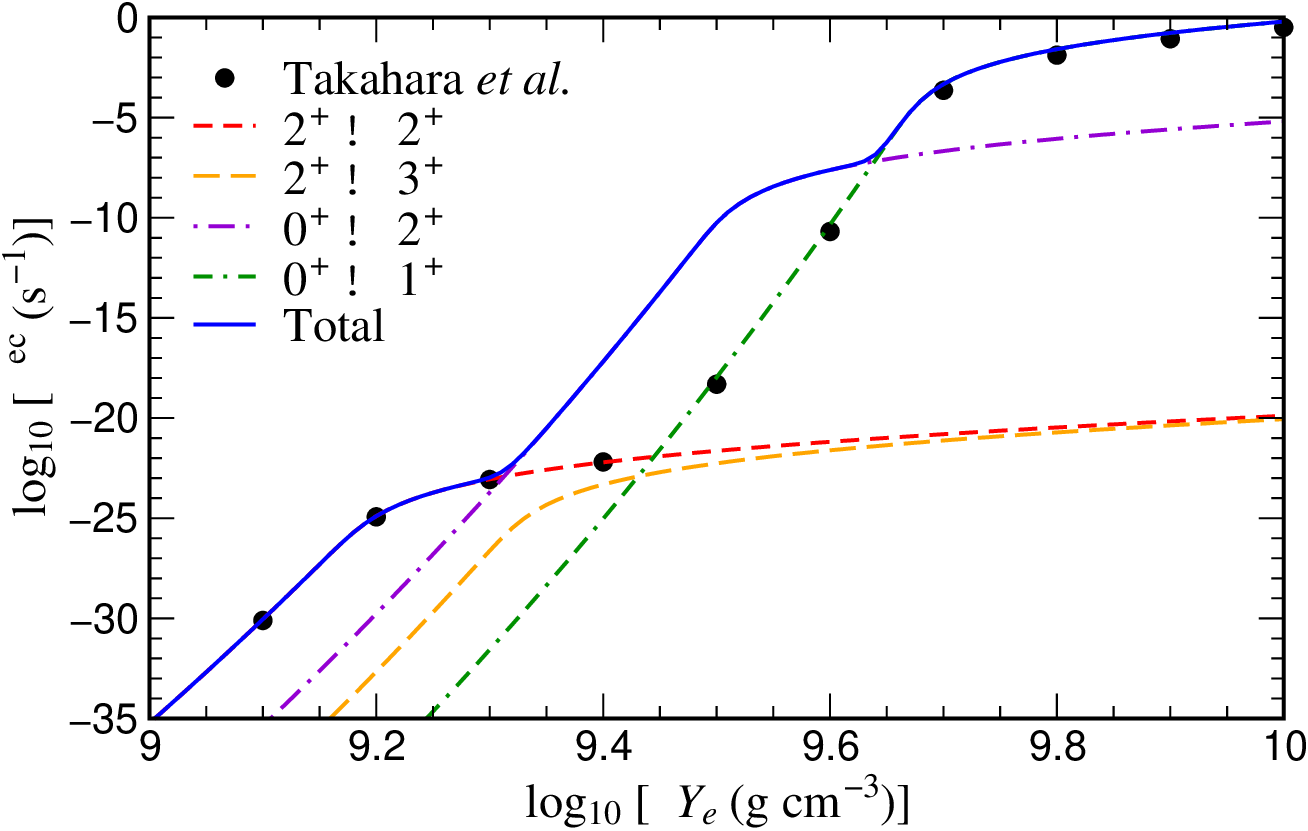}\\
  \includegraphics[width=\linewidth]{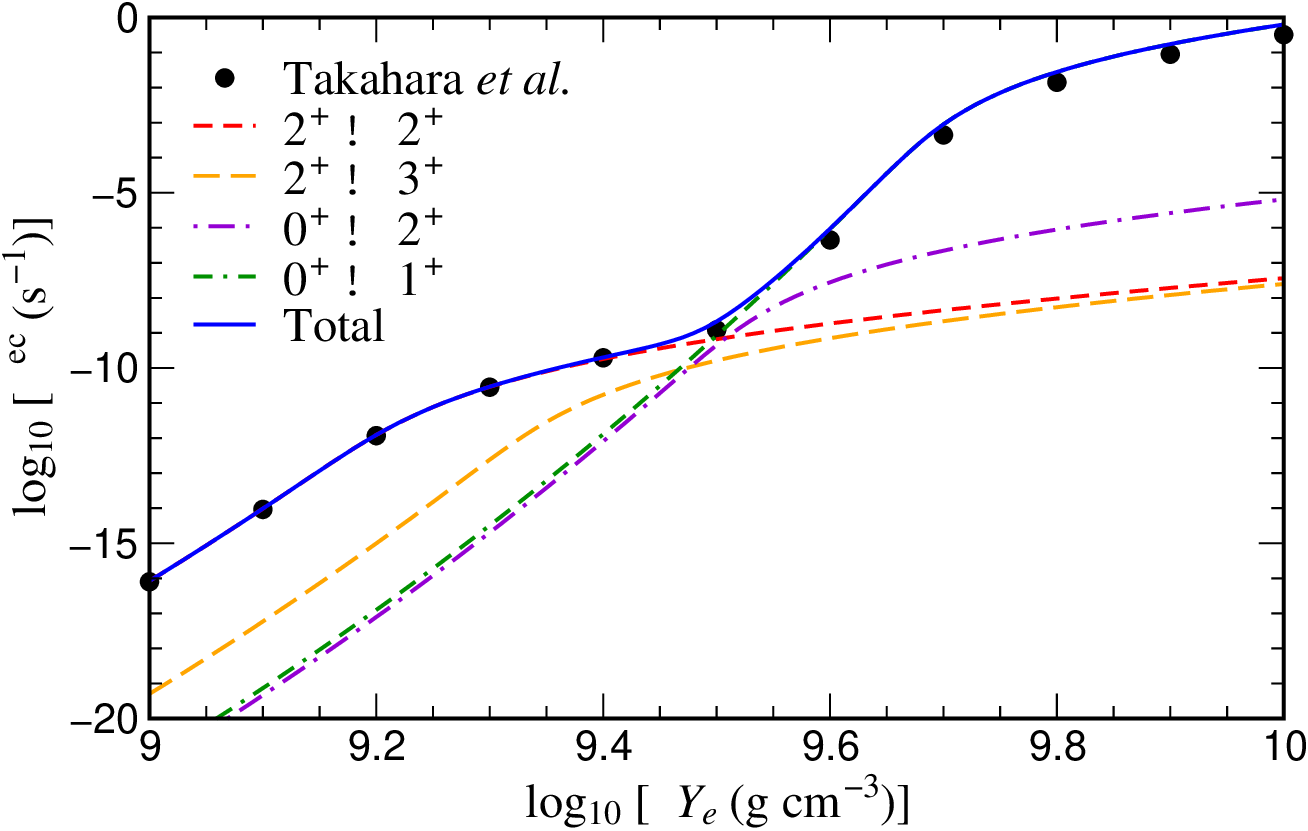}
  \caption{Comparison of our electron capture rate on $^{20}$Ne as
    function of density and for selected temperatures (upper panel:
    $\log\ T(\mathrm{K}) = 8.6$, lower panel: $\log\ T(\mathrm{K}) =
    9.0$) with the values given by Takahara \emph{et al.}
    \cite{Takahara.Hino.ea:1989}. The figure shows the four
    transitions that fully determine the rate. The rates have not been
    corrected for medium effects. \label{fig:ne20ECcomp}}
\end{figure}

\begin{table}[hbt]
  \centering
  \caption{Information that determines the 
    electron capture rate on $^{20}$Ne and beta decay of
    $^{20}$F for the relevant temperatures and
    densities. The ground-state to
    ground-state electron $Q$-value is $Q_{\text{ec}} =
    -7.535$~MeV~\cite{Wang.Audi.ea:2012}. \label{tab:exp20}}   
  \begin{ruledtabular}
    \begin{tabular}{ccccc}
      \multicolumn{2}{c}{Initial $^{20}$Ne state} & \multicolumn{2}{c}{Final $^{20}$F
        state} & Matrix element\\ \cline{1-2}\cline{3-4}
      $J^\pi$ & Energy (MeV)& $J^\pi$ & Energy (MeV) & $B$ \\\hline
      $0^+$ & 0 & $1^+$ & 1.057 & 0.256 \\ 
      $0^+$ & 0 & $2^+$ &  0    & $9.72\times10^{-7}$\footnote{Upper
        experimental limit} \\
      $2^+$ & 1.634 & $2^+$ & 0 & 0.0659\\
      $2^+$ & 1.634 & $3^+$ & 0.656 & 0.0653\footnote{Theoretical value}
    \end{tabular}
  \end{ruledtabular}
\end{table}

In Fig. \ref{fig:ne20ECcomp} we plot our calculated $^{20}$Ne electron
capture rates and compare them to the values presented by Takahara
\emph{et al}~\cite{Takahara.Hino.ea:1989}. The figure shows also the
four transitions that determine the capture rate for the relevant
astrophysical conditions. The difference between the contribution of
these four transitions and the total rate is less than 1\% for the
relevant range of temperatures and densities. The values of the $GT$
matrix elements used are shown in table~\ref{tab:exp20}.

For densities at which the electron chemical potential is smaller than
the electron capture threshold the electron capture rate can be
approximated by equation~\eqref{eq:contijse}, with
$Q=-7.535$~MeV~\cite{Wang.Audi.ea:2012}. Transitions to the final
ground state are favored, i.e. $E_f=0$, which applies for the allowed
transition from the first excited state in $^{20}$Ne and for the
non-unique second forbidden ground-state to ground-state
transition. Due to the larger transition matrix element for the
allowed transition and the lower threshold for capture on the excited
state, the allowed transition from the excited $2^+$ state dominates the
rate as low densities, as can be seen in Fig.~\ref{fig:ne20ECcomp}.

With increasing density the electron chemical potential becomes larger
and the individual electron capture rates initially increase 
exponentially (see discussion following Eq. \eqref{eq:contijse}).
However, once
the electron chemical potential becomes of the order of the threshold
for capture on the $2^+$ excited state of $^{20}$Ne
($E_{\text{thres}} =5.9$~MeV corresponding to $\log\rho Y_e (\textrm{g cm}^{-3}) =
9.2$), the contribution to the $^{20}$Ne electron capture of the $2^+$
state behaves according to equation~\eqref{eq::contijme}; i.e.  
the rate from the $2^+$ grows like a power of the electron
chemical potential and it is suppressed by the Boltzmann factor
$\exp(-E_i/(k T))$, with $E_i = 1.634$~MeV. As the chemical potential
is still lower than the threshold for capture on the ground state to
either the ground state ($E_{\text{thres}} =7.535$~MeV) or
to the first $1^+$ state ($E_{\text{thres}} = 8.592$~MeV) in $^{20}$F the
contributions of these states to the capture rate grow
exponentially and indeed dominate the rate at higher densities 
(see Fig.~\ref{fig:ne20ECcomp}). 
As the threshold for the non-unique second
forbidden transition to the ground state is lower than the one for the
allowed transition to the $1^+$, it can dominate the rate provided
that: 

\begin{equation}
  \label{eq:13}
  \frac{\Lambda^{\text{ec}}(0^+\rightarrow
    2^+)}{\Lambda^{\text{ec}}(0^+\rightarrow 1^+)} =0.77 \frac{ 
    B(0^+ \rightarrow 2^+)}{B(0^+ \rightarrow 1^+)}
  \exp\left(\frac{1.057\ \mathrm{MeV}}{k T}\right) > 1.
\end{equation}

Using the values of the matrix elements from table~\ref{tab:exp20},
the forbidden contribution dominates the rate for temperatures smaller
than 0.9~GK. Figure~\ref{fig:ne20ECcomp} shows that this is in fact
the case, exemplified for the temperature $\log\ T(\text{K}) = 8.6$,
in the density range $\log\rho Y_e (\textrm{g cm}^{-3}) = 9.3$--9.6.

The above results have been obtained assuming an allowed shape for the
phase space of the second forbidden transition. The shape factor for
non-unique second forbidden transitions can contain additional powers
of the electron energy ranging from zero to
four~\cite{Behrens.Buehring:1982}. A dependence like $E_e^2$ will
increase the second forbidden rate by a factor 4, while a dependence
like $E_e^4$ will result in a rate a factor 10 larger, compensating
the possible overestimate of the matrix element by the current
experimental upper limit.

As the second-forbidden transition has not been included in previous
rate estimates~\cite{Takahara.Hino.ea:1989,Oda.Hino.ea:1994}, our rate
is larger in the density regime $\rho Y_e = 2$--$4\times
10^9$~g~cm$^{-3}$. We note that this difference can amount to several
orders of magnitude at temperatures below 0.9 GK. Hence even if the
forbidden transition strength is somewhat smaller than the current
experimental upper limit, this state is likely to dominate the rate in
an important temperature-density range for the evolution of the cores
of 8--12~M$_\odot$ stars.  

At densities beyond $\log \rho Y_e (\textrm{g cm}^{-3}) = 9.6$, the rate is given by the
GT$_+$ transition from the $^{20}$Ne ground state to the lowest $1^+$
state in $^{20}$F. For this transition we use the experimental value
determined from the $(p,n)$ charge-exchange experiment of
Ref.~\cite{Anderson.Tamimi.ea:1991}. This value is in agreement with
the transition strength recently derived from a $(p,p^\prime)$
experiment~\cite{Neumann-Cosel:2013}. In principle the GT strength can
also be obtained from the experimental M1 strength between the
respective states. This is, however, model dependent as this M1
transition has a very strong orbital contribution due to the large
deformation of $^{20}$Ne.  Takahara \emph{et
al.}~\cite{Takahara.Hino.ea:1989} and Oda \emph{et
al.}~\cite{Oda.Hino.ea:1994} had no access to the experimental data and
used a shell model transition strength instead which, however, was a
factor 2 smaller than the experimental value. This explains the
difference between our capture rate at high densities to the previous
calculations.

The most important conclusion from Fig.~\ref{fig:ne20ECcomp} is that
the electron capture rate on $^{20}$Ne is basically fixed by
experimental input, with the exception of the density regime $\log
\rho Y_e (\textrm{g cm}^{-3}) =9.3$--9.6 at temperatures $T < 10^9$ K,
where the forbidden ground-state-to-ground-state transition is likely
to determine the rate. To put the rate entirely on experimental
values, a measurement of this forbidden transition is highly
desirable.

\begin{figure}[htb]
  \begin{center}
    \includegraphics[width=\linewidth]{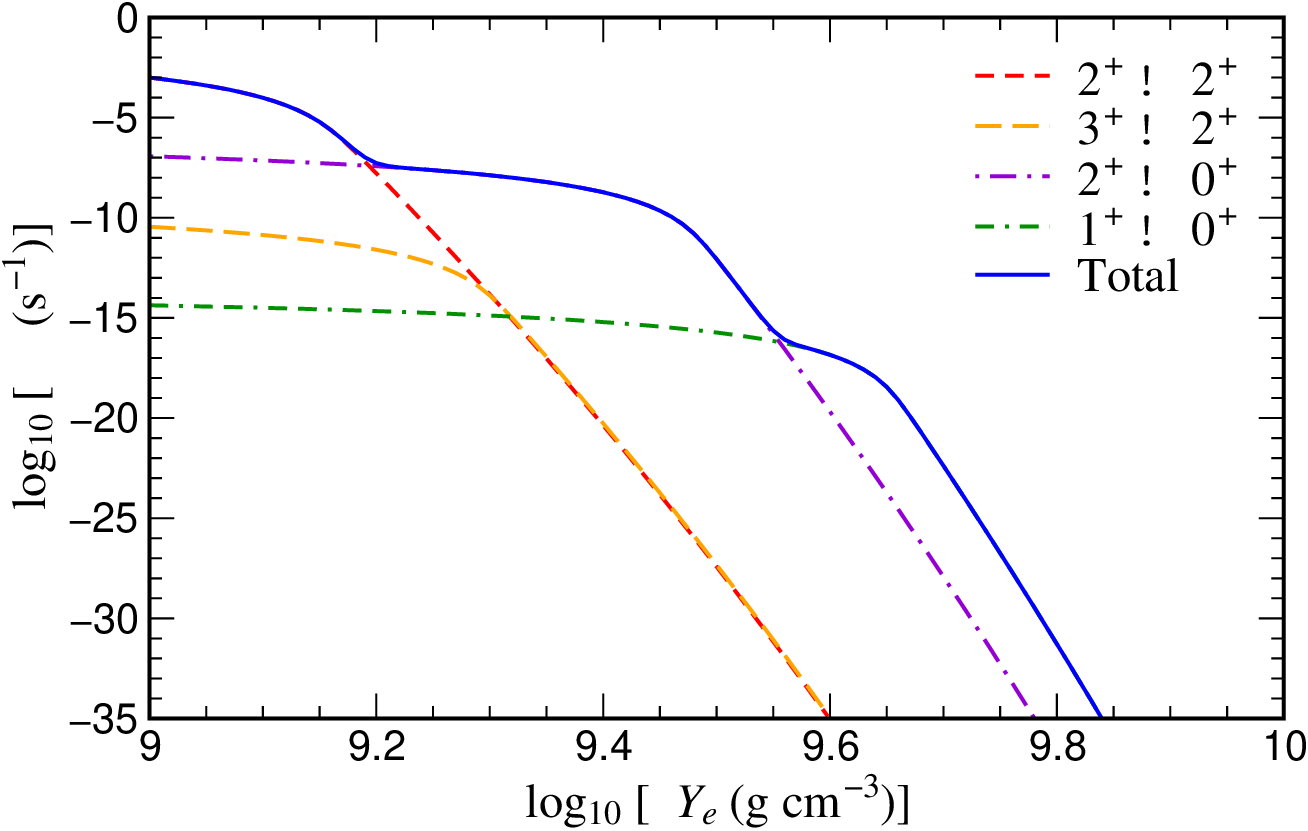}\\
    \includegraphics[width=\linewidth]{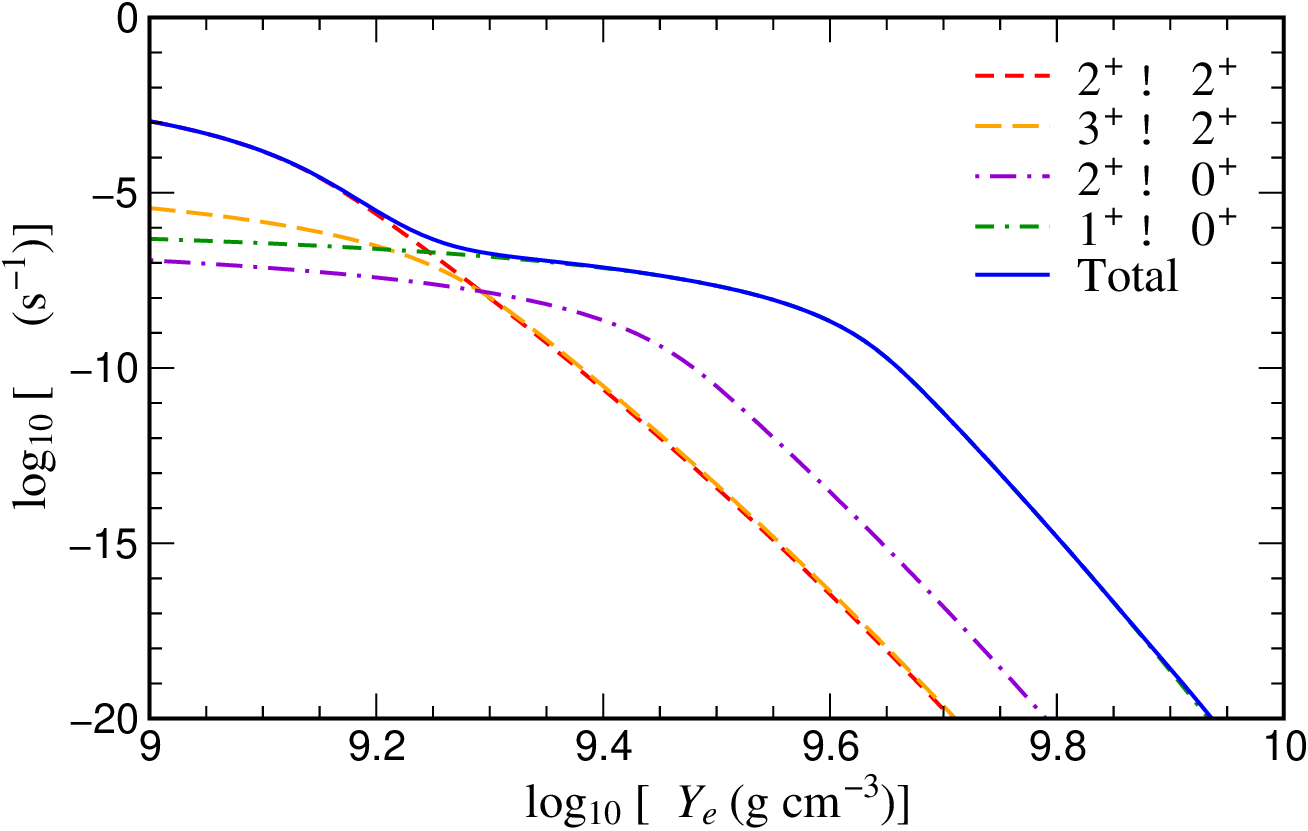}
    \caption{Beta decay rate of $^{20}$F as a function of density and
      for selected temperatures (upper panel: $\log\ T(\mathrm{K}) =
      8.6$, lower panel: $\log\ T(\mathrm{K}) = 9.0$).  The figure
      shows the four transitions that fully determine the rate. The
      rates have not been corrected for medium
      effects.\label{fig:f20betacomp}}
  \end{center}
\end{figure}

\begin{figure}[htb]
  \begin{center}
    \includegraphics[width=\linewidth]{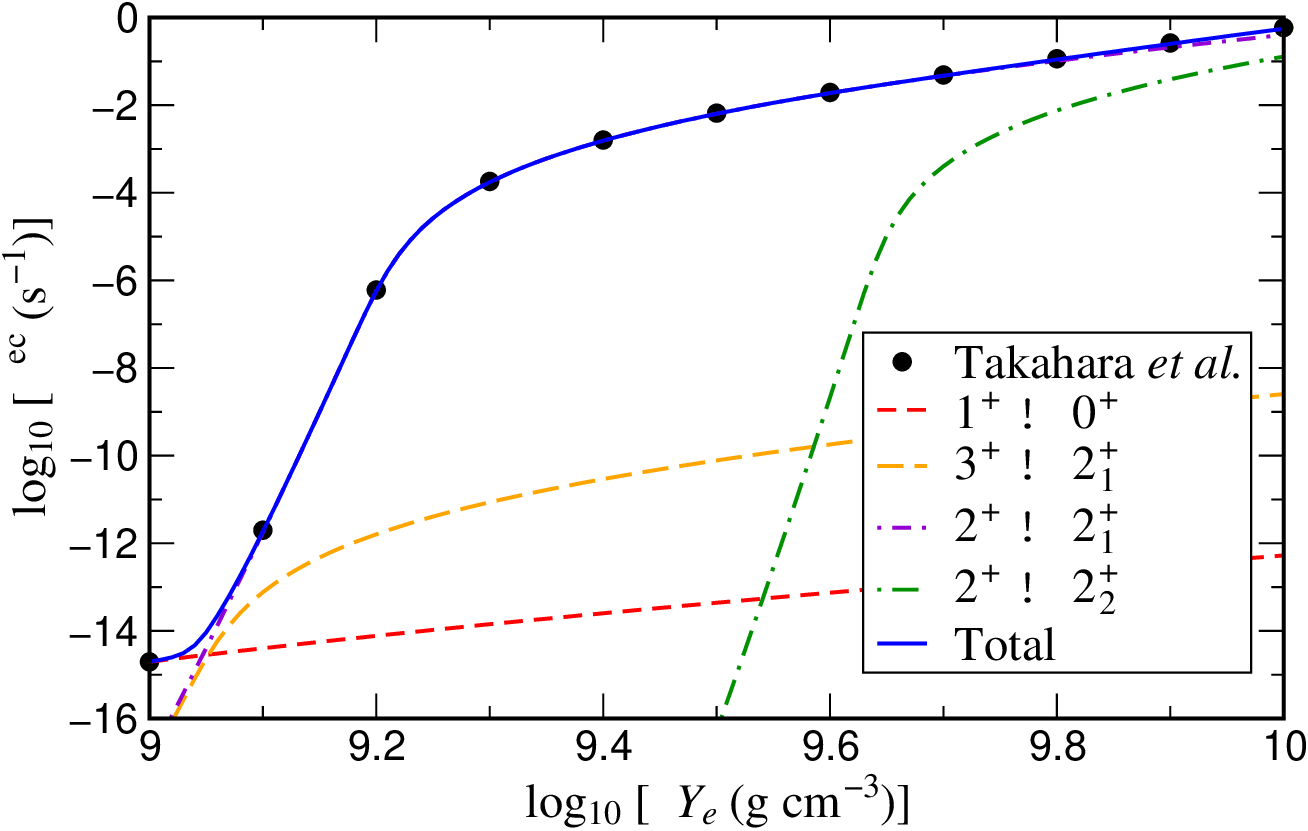}\\
    \includegraphics[width=\linewidth]{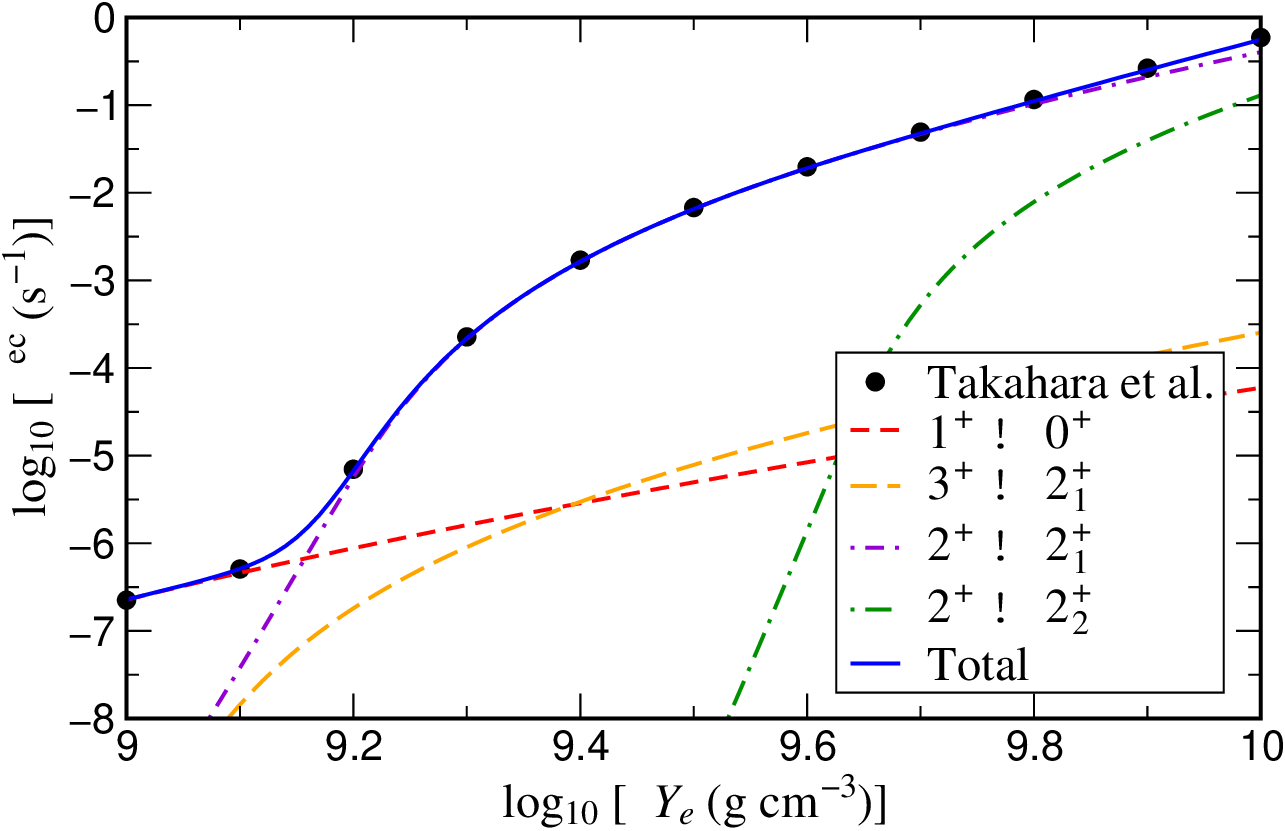}
    \caption{Comparison of our electron capture rate on $^{20}$F as
      function of density and for selected temperatures (upper panel:
      $\log\ T(\mathrm{K}) = 8.6$, lower panel: $\log\ T(\mathrm{K}) =
      9.0$) with the values given by Takahara \emph{et
        al.}~\cite{Takahara.Hino.ea:1989}. The rates have not been 
      corrected for medium effects. \label{fig:f20ECcomp}}
  \end{center}
\end{figure}

The beta-decay rate of $^{20}$F (see Fig.~\ref{fig:f20betacomp})
is determined by the same transitions as the electron capture rate on
$^{20}$Ne. At low densities the decay is dominated by the transition
from the ground state of $^{20}$F to the $2^+$ excited state of
$^{20}$Ne. However, due to the presence of a degenerate electron gas
in the stellar environment
the beta-decay phase space is reduced which increases the decay half-live
with respect to conditions in the laboratory. With increasing density
the electron final state blocking becomes more important and for
densities around $\log\rho Y_e (\textrm{g cm}^{-3}) = 9.2$ the above
transition is fully blocked. At this moment the decay proceeds primarily by
either the ground-state to ground-state second forbidden transition or
by the transition from the first $1^+$ excited state in
$^{20}$Ne. The latter transition has a larger phase space but it is
suppressed at low temperatures by the Boltzmann factor. Similar to
electron capture on $^{20}$Ne the forbidden transition dominates
the beta-decay rate for temperatures $T < 10^9$~K 
in a density regime above $\log\rho Y_e
(\textrm{g cm}^{-3}) = 9.2$. 

\begin{table}[hbt]
  \centering
  \caption{Matrix elements that determine the 
    electron capture rate on $^{20}$F for the relevant temperatures and
    densities. The ground-state to
    ground-state electron $Q$-value is $Q_{\text{ec}} =
    -4.325$~MeV~\cite{Wang.Audi.ea:2012}.\label{tab:expf20}}  
  \begin{ruledtabular}
    \begin{tabular}{ccccc}
      \multicolumn{2}{c}{Initial $^{20}$F state} &
      \multicolumn{2}{c}{Final $^{20}$O 
        state} & Matrix element\\ \cline{1-2}\cline{3-4}
      $J^\pi$ & Energy (MeV)& $J^\pi$ & Energy (MeV) & $B$ \\\hline
      $2^+$ & 0 & $2^+_1$ & 1.674 & 0.0229\footnote{Theoretical value} \\ 
      $2^+$ & 0 & $2^+_2$ & 4.072 & 0.0436\footnotemark[1] \\ 
      $3^+$ & 0.656 & $2^+_1$ & 1.674 & 0.0150\footnotemark[1] \\ 
      $1^+$ & 1.057 & $0^+$ & 0 & 0.378\footnote{From $^{20}$O
        decay~\cite{Tilley.Cheves.ea:1998}}
    \end{tabular}
  \end{ruledtabular}
\end{table}

In Fig.~\ref{fig:f20ECcomp} the electron capture rate on $^{20}$F is
compared with the values computed by Takahara \emph{et
  al}~\cite{Takahara.Hino.ea:1989}.  In the relevant temperature and
density range the rate is mainly dominated by the transition from the
ground state to the $2^+$ excited state in $^{20}$O. The matrix
elements for the transitions determining the rate are given in
table~\ref{tab:expf20}. The theoretical values have been determined by
a shell-model calculation using the USDB
interaction~\cite{Brown.Richter:2006}. At low densities the rate is
determined by the transition from the $1^+$ excited state of $^{20}$F
to the ground state of $^{20}$O which is known experimentally from the
beta decay of $^{20}$O~\cite{Tilley.Cheves.ea:1998}. Our electron
capture rates on $^{20}$F agree with the previous
results~\cite{Oda.Hino.ea:1994,Takahara.Hino.ea:1989}.

The rates shown in figures~\ref{fig:ne20ECcomp},
\ref{fig:f20betacomp} and \ref{fig:f20ECcomp} show clear kinks that
mark the transition between density regimes which are dominated by
different individual transitions. Obviously the kinks get smeared out
with increasing temperature. In astrophysical simulations a reliable
resolution of the kinks requires either a very fine grid for the
tabulation of the rates or an analytical expression. We
will provide this analytical expression in the following. 

For electron capture on $^{20}$Ne we can write the electron capture
rate, based on the 4 transitions identified above, as:
\begin{widetext}
  \begin{equation}
  \label{eq:ne20ecana}
  \begin{split}
  \lambda^{\text{ec}}(^{20}\mathrm{Ne}) = \frac{\ln 2}{K}
  \bigl[&5 e^{-E(2^+)/kT} B_e(2^+\rightarrow 2^+) \Phi^{\text{ec}}_e (Q(2^+\rightarrow 
    2^+),T,\mu_e) 
    + 5 e^{-E(2^+)/kT} B_e(2^+\rightarrow 3^+) \Phi^{\text{ec}}_e (Q(2^+\rightarrow 
    3^+),T,\mu_e)\\
    &+B_e(0^+\rightarrow2^+) \Phi^{\text{ec}}_e (Q(0^+\rightarrow 
    2^+),T,\mu_e)
    +B_e(0^+\rightarrow1^+)\Phi^{\text{ec}}_e(Q(0^+\rightarrow 
    1^+),T,\mu_e)\bigr]    
  \end{split}
\end{equation}
\end{widetext} 

\begin{table}[hbt]
  \centering
  \caption{Numerical values of the quantities that determine the
    analytical expression for the electron capture rate on
    $^{20}$Ne. \label{tab:anaNe20ec}}   
  \begin{ruledtabular}
    \begin{tabular}{ccc}
      Transition & $Q$-value & Effective matrix element \\
      $J^\pi(^{20}\mathrm{Ne})\rightarrow J^\pi(^{20}\mathrm{F})$  &
      (MeV)& $B_e$ \\\hline 
      $2^+\rightarrow 2^+$ & $-5.902$ & 0.0835 \\
      $2^+\rightarrow 3^+$ & $-6.558$ & 0.0827 \\
      $0^+\rightarrow 2^+$ & $-7.536$ & $1.23\times10^{-6}$\\
      $0^+\rightarrow 1^+$ & $-8.592$ & 0.324 \\
    \end{tabular}
  \end{ruledtabular}
\end{table}

The function $\Phi^{\text{ec}}_e$ is defined in
equation~\eqref{eq:ephasefermi}. All other quantities necessary for
the evaluation of the rate are defined in
table~\ref{tab:anaNe20ec}. The difference between the effective
transition matrix element, $B_e$, appearing in table
\ref{tab:anaNe20ec} and the matrix element, $B$, of
table~\ref{tab:exp20} is due to the fact that the former includes the
average value of the Fermi Coulomb distortion function. This value is
equal to 1.267 for $Z=10$.  Equation~(\ref{eq:ne20ecana}) reproduces
the electron capture rate on $^{20}$Ne if one uses ``exact'' numerical
values for the Fermi integrals. Using the approximate expressions of
the Fermi function defined in eq.~\eqref{eq:fermiapprox}, the largest
error is around 15\%.  This uncertainty occurs when the electron fermi
energy is of the order of the $Q$-value of the dominating
transition. The origin of the uncertainty is mainly due to the
approximation made in the Fermi integral of order 2 in
equation~\eqref{eq:fermi2}.

The beta-decay rate of $^{20}$F is given by the same 4 transitions
as the electron capture on $^{20}$Ne for the astrophysically
relevant conditions of interest here. The rate can then be written as: 
\begin{widetext}
\begin{equation}
  \label{eq:f20beana}
  \begin{split}
  \lambda^{\beta}(^{20}\mathrm{F}) = \frac{\ln 2}{5 K}
  \bigl[&5 B_e(2^+\rightarrow 2^+) \Phi^{\beta}_e (Q(2^+\rightarrow 
    2^+),T,\mu_e) + 7 e^{-E(3^+)/kT} B_e(3^+\rightarrow 2^+)
    \Phi^{\beta}_e (Q(3^+\rightarrow  2^+),T,\mu_e)\\
    &+5 B_e(2^+\rightarrow 0^+) \Phi^{\beta}_e (Q(2^+\rightarrow 
    0^+),T,\mu_e)
    +3 e^{-E(1^+)/kT} B_e(1^+\rightarrow 0^+)\Phi^{\beta}_e(Q(1^+\rightarrow 
    0^+),T,\mu_e)\bigr]
  \end{split}
\end{equation}
\end{widetext}

\begin{table}[hbt]
  \centering
  \caption{Numerical values of the quantities that determine the
    analytical expression for the beta decay rate of
    $^{20}$F. \label{tab:anaF20B}}   
  \begin{ruledtabular}
    \begin{tabular}{ccc}
      Transition & $Q$-value & Effective matrix element \\
      $J^\pi(^{20}\mathrm{F})\rightarrow J^\pi(^{20}\mathrm{Ne})$  &
      (MeV)& $B_e$ \\\hline 
      $2^+\rightarrow 2^+$ & $5.902$ & 0.0835 \\
      $3^+\rightarrow 2^+$ & $6.558$ & 0.0591 \\
      $2^+\rightarrow 0^+$ & $7.536$ & $2.46\times10^{-7}$\\
      $1^+\rightarrow 0^+$ & $8.592$ & 0.108 \\
    \end{tabular}
  \end{ruledtabular}
\end{table}

The function $\Phi^{\beta}_e$ is defined in
equation~\eqref{eq:betaphasefermiapp}. All other quantities necessary
for the evaluation of the rate are defined in
table~\ref{tab:anaF20B}. Similarly to the case of electron
capture on $^{20}$Ne the rate is reproduced ``exactly'' if the Fermi
integrals are evaluated to numerical precision and errors of around
15\% are obtained if one uses the approximate expressions defined in
eq.~\eqref{eq:fermiapprox}.

Finally, the electron capture rate on $^{20}$F can be written as: 

\begin{widetext}
\begin{equation}
  \label{eq:f20ecana}
  \begin{split}
  \lambda^{\text{ec}}(^{20}\mathrm{F}) = \frac{\ln 2}{5 K}
  \bigl[&3 e^{-E(1^+)/kT} B_e(1^+\rightarrow 0^+) \Phi^{\text{ec}}_e
  (Q(1^+\rightarrow  0^+),T,\mu_e) + 7 e^{-E(3^+)/kT}
  B_e(3^+\rightarrow 2^+_1) \Phi^{\text{ec}}_e (Q(3^+\rightarrow
  2^+_1),T,\mu_e)\\  
    &+ 5 B_e(2^+\rightarrow 2^+_1) \Phi^{\text{ec}}_e (Q(2^+\rightarrow
    2^+_1),T,\mu_e) + 5 B_e(2^+\rightarrow 2^+_2) \Phi^{\text{ec}}_e
    (Q(2^+\rightarrow  2^+_2),T,\mu_e)\bigr]
  \end{split}
\end{equation}
\end{widetext}

\begin{table}[hbt]
  \centering
  \caption{Numerical values of the quantities that determine the
    analytical expression for the electron capture on
    $^{20}$F. \label{tab:anaF20ec}}   
  \begin{ruledtabular}
    \begin{tabular}{ccc}
      Transition & $Q$-value & Effective matrix element \\
      $J^\pi(^{20}\mathrm{F})\rightarrow J^\pi(^{20}\mathrm{Ne})$  &
      (MeV)& $B_e$ \\\hline 
      $1^+\rightarrow 0^+$ & $-3.268$ & 0.467 \\
      $3^+\rightarrow 2^+_1$ & $-5.342$ & 0.0185 \\
      $2^+\rightarrow 2^+_1$ & $-5.998$ & 0.0283 \\
      $2^+\rightarrow 2^+_2$ & $-8.397$ & 0.0539 
    \end{tabular}
  \end{ruledtabular}
\end{table}

The quantities necessary for the evaluation of the rate are defined in
table~\ref{tab:anaF20ec}. The difference between the effective
transition matrix element, $B_e$, appearing in
table~\ref{tab:anaF20ec} and the matrix element, $B$, of
table~\ref{tab:expf20} is due to the fact that the former includes the
average value of the Fermi Coulomb distortion function. This value is
equal to 1.236 for $Z=9$. 

The above expressions can be generalized to the calculation of the
neutrino energy loss rate by simply substituting the function $\Phi$
by $\Psi$ defined in equations~\eqref{eq:lossfermi}
and~\eqref{eq:betaephasefermiapp} for electron capture and beta-decay,
respectively.

Coulomb corrections are an important modification of weak interaction
rates in dense astrophysical environment. In the following we include
screening corrections in our rates following the generalization of the
screening treatment, as originally developed by Bravo and Garcia-Senz
\cite{Bravo.Garcia-Senz:1999}, and presented in detail for electron
capture in the appendix of Ref.
\cite{Juodagalvis.Langanke.ea:2010}. Screening has two effects on
electron capture: the threshold energy in the medium is increased,
$Q_{if}^{\text{ec,med}} = Q^{\text{ec}}_{if} - \Delta Q_c(Z)$ (notice
that $Q_{if}$ is negative in our convention), and the chemical
potential of the electrons is reduced, $\mu_e^{\text{med}} = \mu_e - V_s$,
where the parameters $\Delta Q_c(Z)$ and $V_s$ can be calculated
following~\cite{Juodagalvis.Langanke.ea:2010,Bravo.Garcia-Senz:1999,%
  Itoh.Tomizawa.ea:2002}. We note that both effects reduce the
electron capture rate. The opposite is true for $\beta^-$ decays where
screening enhances the rate. At first, the lowering of the electron
chemical potential results in reduction of the Pauli blocking in the
final state, i.e. smaller values of ($1-S_e(\omega)$) in
Eq.~\eqref{eq:brateb}.  Secondly, for beta decays the change in
threshold is given by $Q_{if}^{\beta,\text{med}} = Q^{\beta}_{if} + \Delta
Q_c (Z+1))$ (note that $Q^{\beta}_{if} = - Q^{\text{ec}}_{fi}$).

\begin{figure}[htb]
  \begin{center}
    \includegraphics[width=\linewidth]{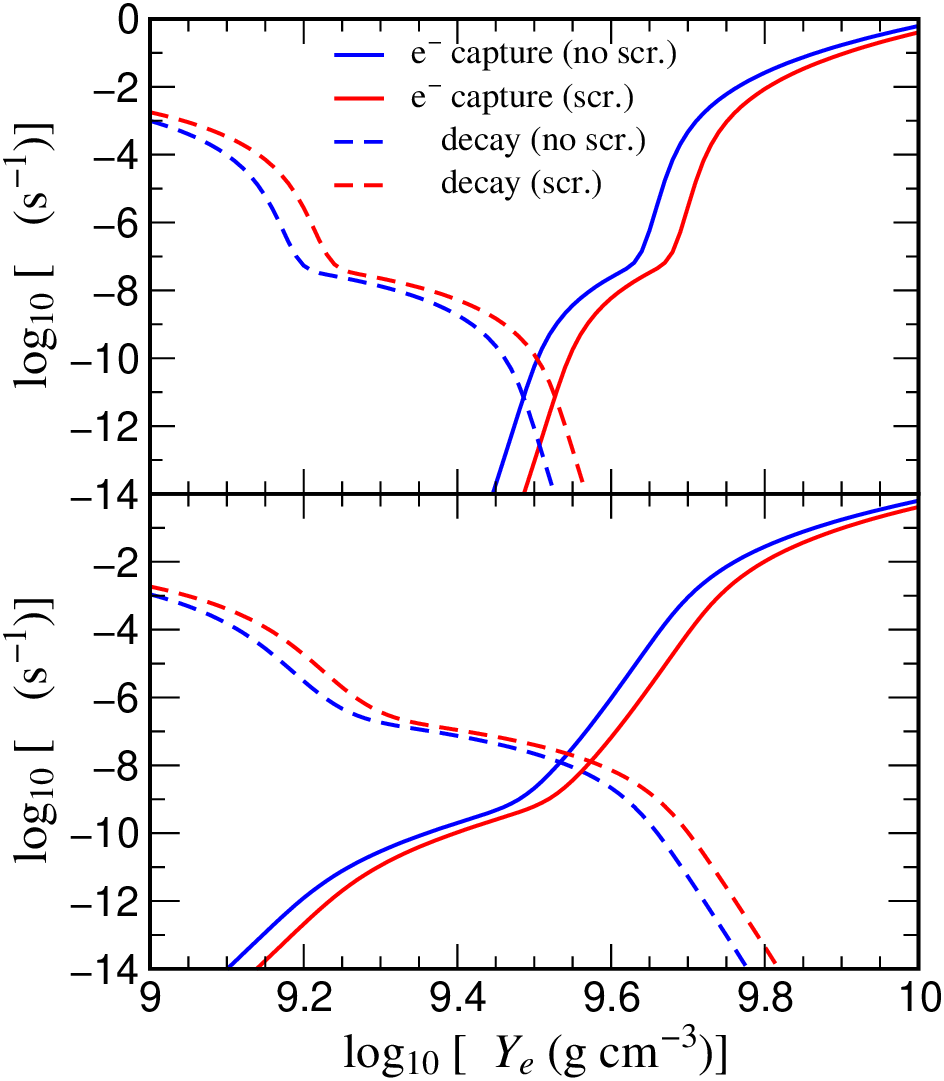}
    \caption{Electron capture rate on $^{20}$Ne and beta decay rate of
      $^{20}$F with and without consideration of medium corrections
      (upper panel: $\log\ T(\mathrm{K}) = 8.6$, lower panel: $\log\
      T(\mathrm{K}) = 9.0$).\label{fig:ne20f-scr}}
  \end{center}
\end{figure}

To demonstrate the effect of screening we show in
figure~\ref{fig:ne20f-scr} the electron capture rate on $^{20}$Ne and
the beta-decay rate of $^{20}$F with and without consideration of
Coulomb corrections. Obviously the effect is largest at low
temperatures as both rates are more sensitive to modifications of the
threshold energy which changes by 120~keV at the lowest  and by
270~keV at the highest density considered in
Fig.~\ref{fig:ne20f-scr}.
More importantly, screening corrections change
the density at which electron capture dominates over beta-decay and
can potentially affect the evolution of the
star~\cite{Gutierrez.Garcia-Berro.ea:1996}.
In particular the Coulomb modifications shift the densities at which
URCA pairs operate in stars to higher densities.

\begin{figure}[htb]
  \centering
  \includegraphics[width=\linewidth]{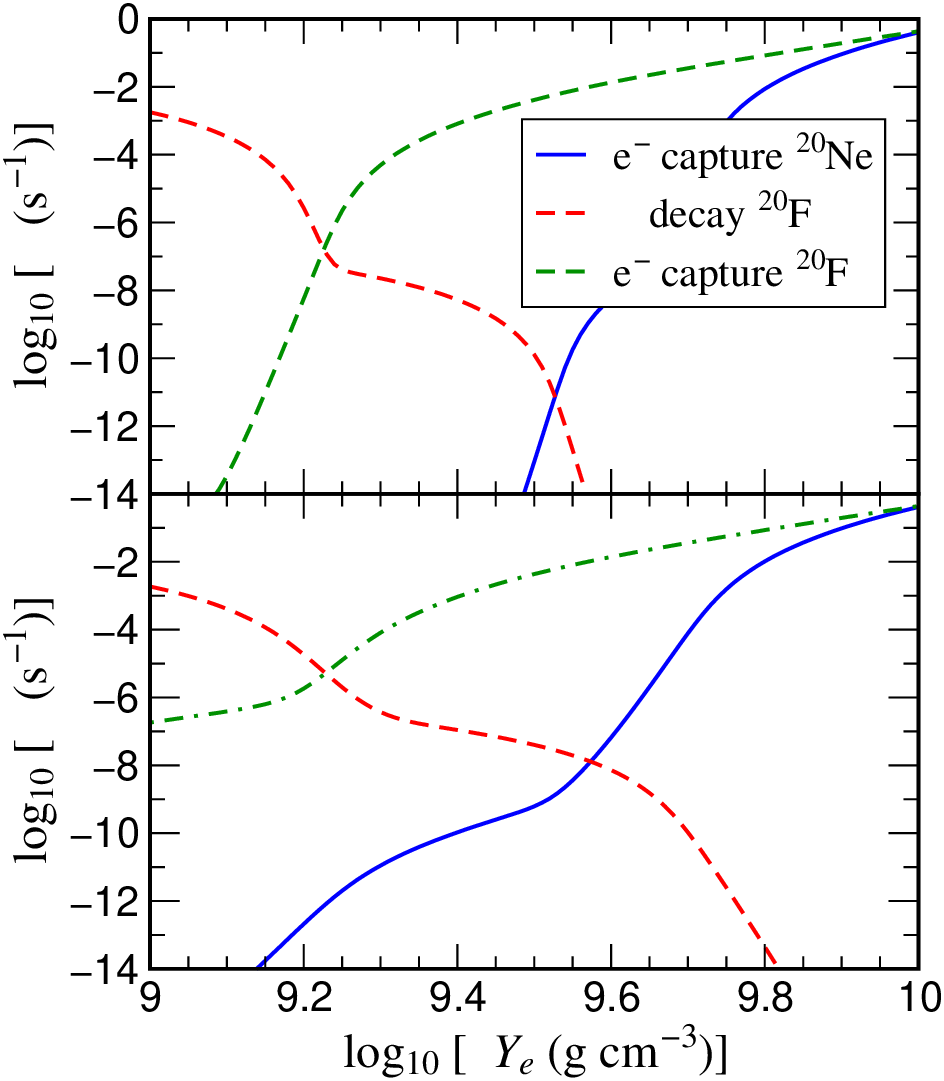}
  \caption{Rates for electron captures on $^{20}$Ne and $^{20}$F and
    beta-decay of $^{20}$F for selected temperatures: $\log\
    T(\mathrm{K}) = 8.6$ (upper panel), $\log\ T(\mathrm{K}) = 9.0$
    (lower panel). Medium corrections are included in the
    rates. \label{fig:ne20f20}} 
\end{figure}

In Fig.~\ref{fig:ne20f20} we compare the $^{20}$Ne electron capture
rate with the rates for the competing $^{20}$F $\beta$ decay and
electron capture, considering Coulomb corrections
as discussed above. We observe that the electron capture rate gets
larger than the competing $^{20}$F decay rate for densities larger
than $\rho Y_e \approx 4 \times 10^9$~g~cm$^{-3}$. As the electron
capture rate on $^{20}$F is faster than the one on $^{20}$Ne, caused
by the smaller $Q$-value, the capture on $^{20}$Ne is followed by a
second capture process leading to $^{20}$O.

\begin{figure}[htb]
  \centering
  \includegraphics[width=\linewidth]{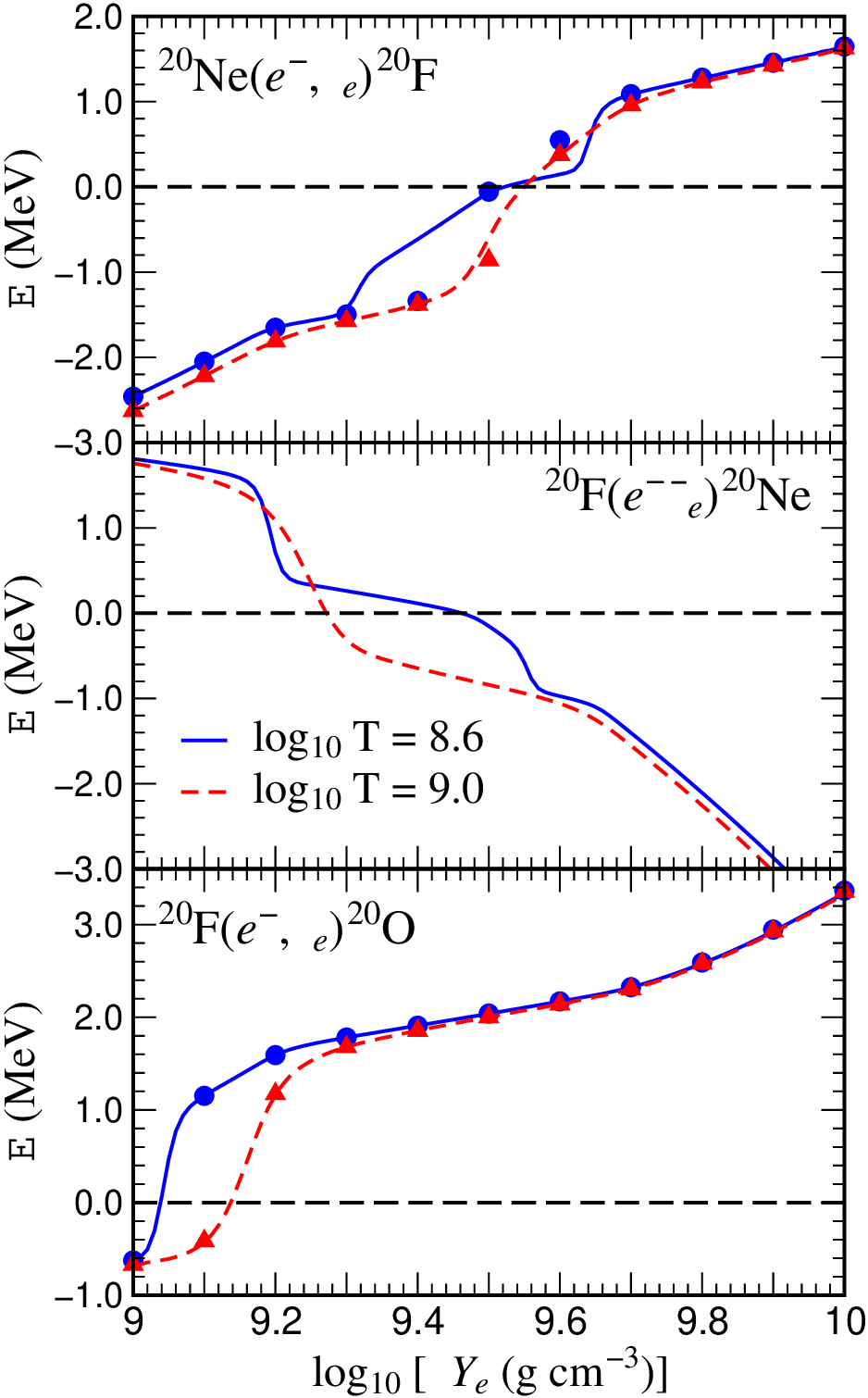}
  \caption{Average energy produced (positive) or absorbed (negative)
    by electron capture on $^{20}$Ne (upper panel), beta-decay of
    $^{20}$F (middle panel) and electron capture on
    $^{20}$F (lower panel).\label{fig:energe20}} 
\end{figure}

At the densities of concern here the neutrinos produced in the
electron captures and $\beta$ decays leave the star unhindered and
carry away energy.  Depending on the amount of energy carried away,
weak processes can result in a net heating or cooling of the stellar
environment. This depends on the sign of the quantity $\mathcal{E}$
defined in equations~\eqref{eq:egeneec}
and~\eqref{eq:egenebeta}. Figure~\ref{fig:energe20} shows
$\mathcal{E}$, i.e. the average energy produced or absorbed by the
various processes (electron capture on $^{20}$Ne, beta-decay of $^{20}$F and
electron capture on $^{20}$F) compared with the values provided by
Takahara \emph{et al.}~\cite{Takahara.Hino.ea:1989}. As discussed in
section~\ref{sec:determ-energy-gener} at low densities, electron
capture is endothermic as energy has to be absorbed from the medium to
populate the excited states that dominate the rate. Under these
conditions the average energy grows proportional to $\mu_e \approx
\rho^{1/3}$ until the rate is dominated by a different
transition. At this moment a sudden change in the growth of the energy
generation takes place. For $^{20}$Ne this transition occurs for
densities $\log\rho Y_e (\textrm{g cm}^{-3}) \approx 9.3$ and
temperatures smaller than 1~GK, when the rate is dominated by the
second-forbidden transition to the $^{20}$F ground state for both the
electron capture on $^{20}$Ne and the beta-decay of $^{20}$F. As this
transition was not included in the work of Takahara \emph{et al}~\cite{Takahara.Hino.ea:1989}, we
predict substantially different values for the energy generation. In the case of
beta-decay, whenever the energy generation is dominated by a
particular transition it decreases like $\mu_e/4$ (see
section~\ref{sec:determ-energy-gener}) and it is positive up to
densities for which $\mu_e$ equals the ground-state to ground-state
$Q$-value. At these densities, the $^{20}$Ne electron capture rate energy
generation becomes positive and changes its growth behavior to
$\mu_e/4$ while beta-decay, being Pauli blocked, decreases like
$-\mu_e$. Due to the smaller $Q$-value for electron capture on
$^{20}$F the average energy produced is positive for most of the
density region shown in figure~\ref{fig:energe20}. For densities
larger than $\log\rho Y_e (\textrm{g cm}^{-3}) \approx 9.2$ the energy
produced by electron capture on $^{20}$F dominates over the energy
loss by capture on $^{20}$Ne and makes the net energy generation
positive. This density marks the transition at which
the net effect of the sequence of weak-interaction processes changes from
endothermic to exothermic and, correspondingly, the core temparture increases 
in stellar evolution
models~\cite{Jones.Hirschi.ea:2013}. The increase in the energy
generation by electron capture on $^{20}$F at high densities is due to
the contribution of transitions to excited states on $^{20}$O (see
figure~\ref{fig:f20ECcomp}) that
increase the average gamma energy in equation~\eqref{eq:energenec}. 
We note that the transition at which electron capture on $^{20}$Ne becomes
exothermic occurs at slightly higher densities than for the capture on
$^{20}$F. This is due to pairing effects which makes the $Q$ value
for the transition from an even-even nucleus to an odd-odd nucleus
larger than the neighboring one from an odd-odd to an even-even nucleus.

\subsection{Rates for the $A=24$ nuclei}

Since the evaluation of the electron capture rate on $^{24}$Mg by
Takahara \emph{et al.}~\cite{Takahara.Hino.ea:1989} and by Oda
\emph{et al.}~\cite{Oda.Hino.ea:1994}, several important experimental
data sets became available. These include an improved measurement of
the $\beta$ decay of the $1^+$ isomeric state in $^{24}$Al which,
assuming isospin symmetry, determines the GT transition from the
analogue state in $^{24}$Na at $E_x=0.426$~MeV to the $^{24}$Mg ground
state and the two excited $2^+$ states at $E_x=1.369$~MeV and 4.238~MeV~\cite{Nishimura.Fujita.ea:2011}.  The GT strength has also been
measured by $(p,n)$ \cite{Anderson.Tamimi.ea:1991},
$(^{3}\mathrm{He},t)$~\cite{Zegers.Meharchand.ea:2008},
$(t,{}^{3}\mathrm{He})$~\cite{Howard.Zegers.ea:2008} and
$(d,{}^2\mathrm{He})$~\cite{Rakers.Baeumer.ea:2002} charge-exchange
reactions where the first two can be applied in the case of $^{24}$Mg
due to the isospin symmetry of the nucleus. The various measurements
basically agree on the $B(GT)$ value for the transition from the $1^+$
isomeric state to the ground state which, as we will see below,
determines the electron capture rate on $^{24}$Mg for a large range of
the astrophysically relevant temperatures and densities.  In the
following we will adopt the value $B(GT)=0.094(3)$, derived from the
$\beta^+$ decay, for this transition. It is slightly larger than the
valures determined from the charge-exchange experiments
($B(GT)=0.079(2)$ from $(p,n)$ \cite{Anderson.Tamimi.ea:1991},
$B(GT)=0.086(2)$ from
$(^{3}\mathrm{He},t)$~\cite{Zegers.Meharchand.ea:2008}
$0.078(8)\pm0.04$ from
$(d,^2\mathrm{He})$~\cite{Rakers.Baeumer.ea:2002}), except the
strength determined by
$(t,^{3}\mathrm{He})$~\cite{Howard.Zegers.ea:2008}, $B(GT)=0.13(2)$
that cannot be separated from the nearby $2^+$ state. However, the
value adopted by us is noticeably larger than the $B(GT)$ value used in the previous
electron capture rate evaluations
\cite{Takahara.Hino.ea:1989,Oda.Hino.ea:1994} derived from $\beta$
decay data available at the time these works were performed.

In details, our input for the calculation of the $^{24}$Mg electron
capture rate is based on the $^{24}$Al $\beta^+$ decay data supplying
the $GT$ transitions from the ground state and the lowest two $2^+$
states in $^{24}$Mg to the $1^+$ state in $^{24}$Na at
$E_x=472$~keV. From the $^{24}$Na $\beta^-$ decay of the $4^+$ ground
state we adopt the $GT$ transitions from the first $4^+$ state in
$^{24}$Mg at $E_x=4.123$~MeV and the second-forbidden transition from
the $2^+$ state at $E_x=1.369$~MeV. The GT values from the ground
state to the other excited $1^+$ states are taken from the
$(^{3}\mathrm{He},t)$ experiment of
Ref.~\cite{Zegers.Meharchand.ea:2008}. Finally, we supplement these
data by GT strength distributions for excited states in $^{24}$Mg
derived from shell model calculations performed in the sd shell and
using the USDB interaction. We note that, due to the strong angular
momentum mismatch, forbidden transitions to the $^{24}$Na $4^+$ ground
state do not contribute to the capture rate. The energies and
$B(GT)$ values which determine the $^{24}$Mg electron capture  
and $\beta$-decay rates at the conditions of interest here are
summarized in Table \ref{tab:exp24}.

\begin{table}[hbt]
  \centering
  \caption{Information that determines the 
    electron capture rate on $^{24}$Mg and beta decay of
    $^{24}$Na for the relevant temperatures and
    densities. The ground-state to
    ground-state electron $Q$-value is $Q_{\text{ec}} =
    -6.026$~MeV~\cite{Wang.Audi.ea:2012}. \label{tab:exp24}}   
  \begin{ruledtabular}
    \begin{tabular}{ccccc}
      \multicolumn{2}{c}{Initial $^{24}$Mg state} & \multicolumn{2}{c}{Final $^{24}$Na
        state} & Matrix element\\ \cline{1-2}\cline{3-4}
      $J^\pi$ & Energy (MeV)& $J^\pi$ & Energy (MeV) & $B$ \\\hline
      $0^+$ & 0 & $1^+_1$ & 0.472 & 0.094(3)\footnote{From decay of mirror
        $^{24}$Al~\cite{Nishimura.Fujita.ea:2011}}\\ 
      $0^+$ & 0 & $1^+_2$ & 1.347 & 1.038(87)\footnote{From $(d,^2\text{He})$~\cite{Rakers.Baeumer.ea:2002}}\\
      $0^+$ & 0 & $1^+_3$ & 1.89 & 0.040(13)\footnotemark[2]\\
      $0^+$ & 0 & $1^+_4$ & 3.41 & 0.460(38)\footnotemark[2]\\
      $2^+$ & 1.369 & $4^+$ &  0.0 & $5.0(3)\times
      10^{-8}$\footnote{From $^{24}$Na decay~\cite{Firestone:2007}}\\
      $2^+$ & 1.369 & $1^+$ &  0.472 & 0.0046(11)\footnotemark[1] \\
      $2^+$ & 1.369 & $2^+$ & 0.563 & 0.032\footnote{Theoretical value}\\
    \end{tabular}
  \end{ruledtabular}
\end{table}

\begin{figure}[htb]
  \centering
  \includegraphics[width=\linewidth]{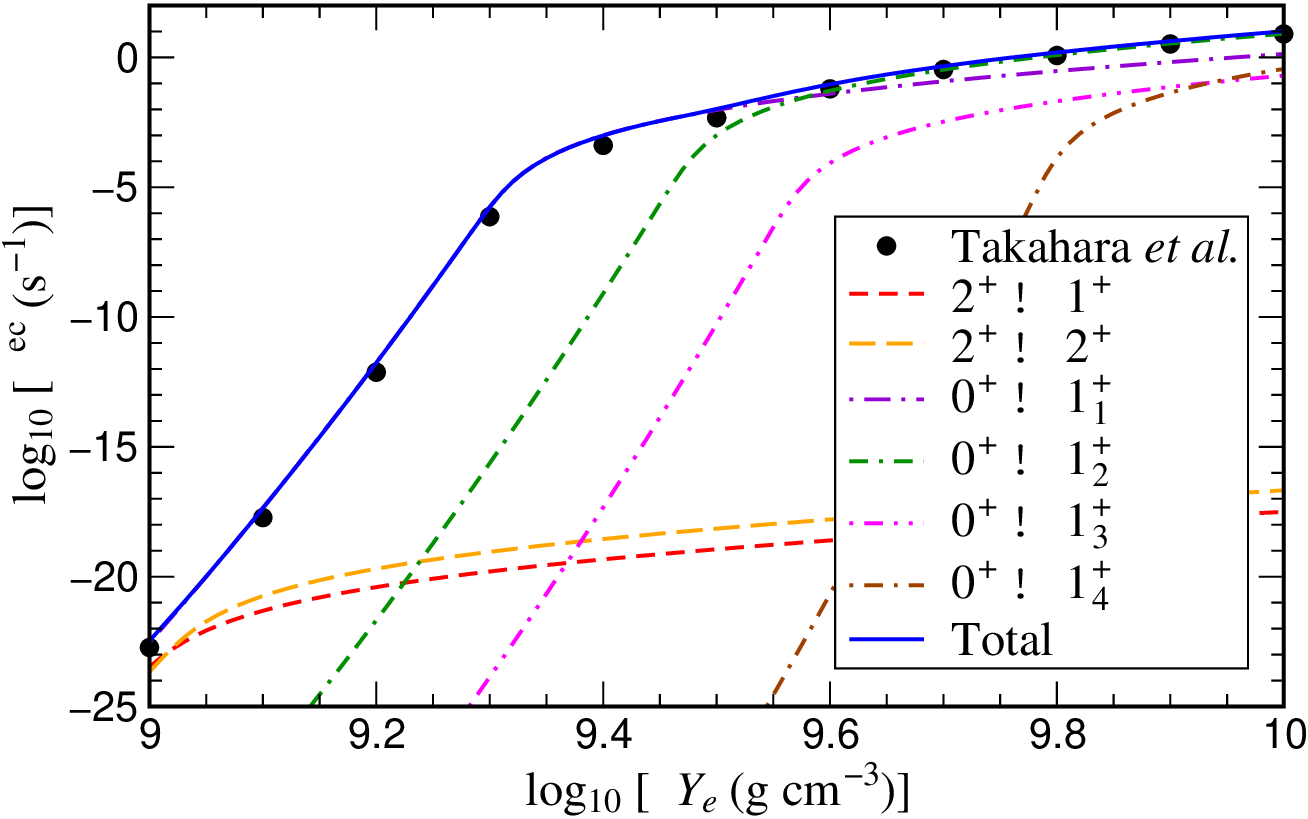}\\
  \includegraphics[width=\linewidth]{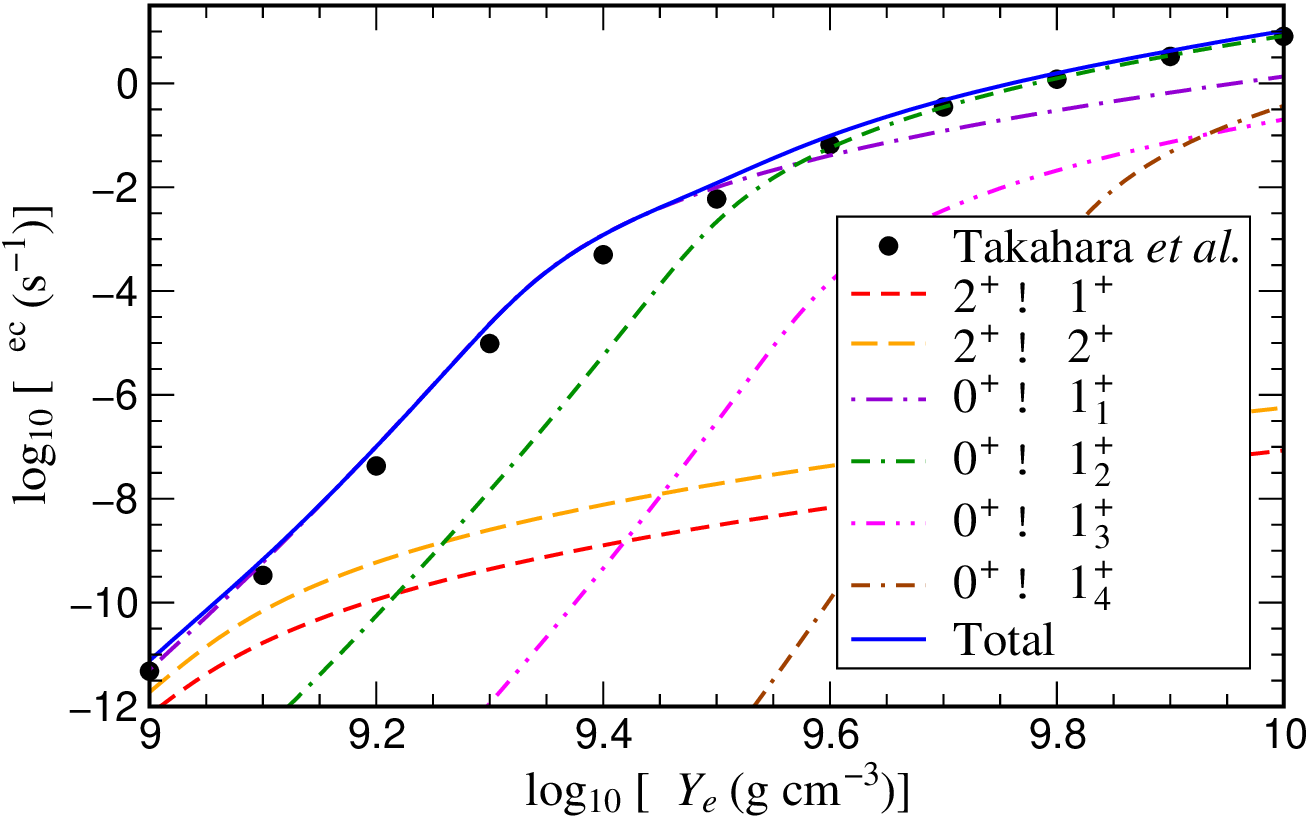}
  \caption{Comparison of our electron capture rate on $^{24}$Mg as
    function of density and for selected temperatures (upper panel:
    $\log\ T(\mathrm{K}) = 8.6$, lower panel: $\log\ T(\mathrm{K}) =
    9.0$) with the values given by Takahara \emph{et
      al.}~\cite{Takahara.Hino.ea:1989}. The figure shows the six
    transitions that fully determine the rate.  The rates have not
    been corrected for medium effects. \label{fig:mg24-capture}}
\end{figure}

Fig. \ref{fig:mg24-capture} shows the electron capture rate on
$^{24}$Mg in the astrophysically relevant density range and for
selected temperatures, not considering medium-induced Coulomb
corrections. For densities $9.0 < \log \rho Y_e (\textrm{g cm}^{-3}) <
9.5$ and the relevant temperature regime ($T=0.4-1.0$) GK the rate is
dominated by the capture from the $^{24}$Mg ground state to the
isomeric state in $^{24}$Na at $E_x=0.472$~MeV. At the highest
temperatures (the lower panel of Fig. \ref{fig:mg24-capture} shows the
rate for $T=10^9$ K) the excited state at $E_x=1.369$~MeV gets
sufficiently thermally populated that its transitions to the excited
$1^+$ state at $E_x=0.472$~MeV and $2^+$ state at $E_x=0.563$ MeV
slighty contribute to the rate at densities $\log \rho Y_e (\textrm{g
  cm}^{-3}) < 9.1$.  At higher densities $\log \rho Y_e (\textrm{g
  cm}^{-3}) > 9.5$ the most important contribution to the rate comes
from the strong GT transition from the ground state to the $1^+$ state
at $E_x=1.347$ MeV ($B(GT)=1.038(87)$~\cite{Rakers.Baeumer.ea:2002})
which is more than 10 times larger than the one to the state at
$E_x=0.472$~MeV and compensates for the larger phase space factor of
the latter at the higher densities. Due to its large B(GT) value of
0.46, the transition from the ground state to the $1^+$ state at
$E_x=3.41$ MeV contributes on the few percent level at $\log \rho Y_e
(\textrm{g cm}^{-3}) 
=10$. The electron capture rate on $^{24}$Mg presented here is
entirely based on experimental input, except for the small
contributions arising from the ground state transitions to the lowest
excited $1^+$ and $2^+$ states, for which we adopt the $B(GT)$ values
from a shell model claculation. These transitions, however, modify the
rate only at low densities ($\log \rho Y_e (\textrm{g cm}^{-3}) < 9.1$) and the highest
temperatures of interest.

\begin{figure}[htb]
  \centering
  \includegraphics[width=\linewidth]{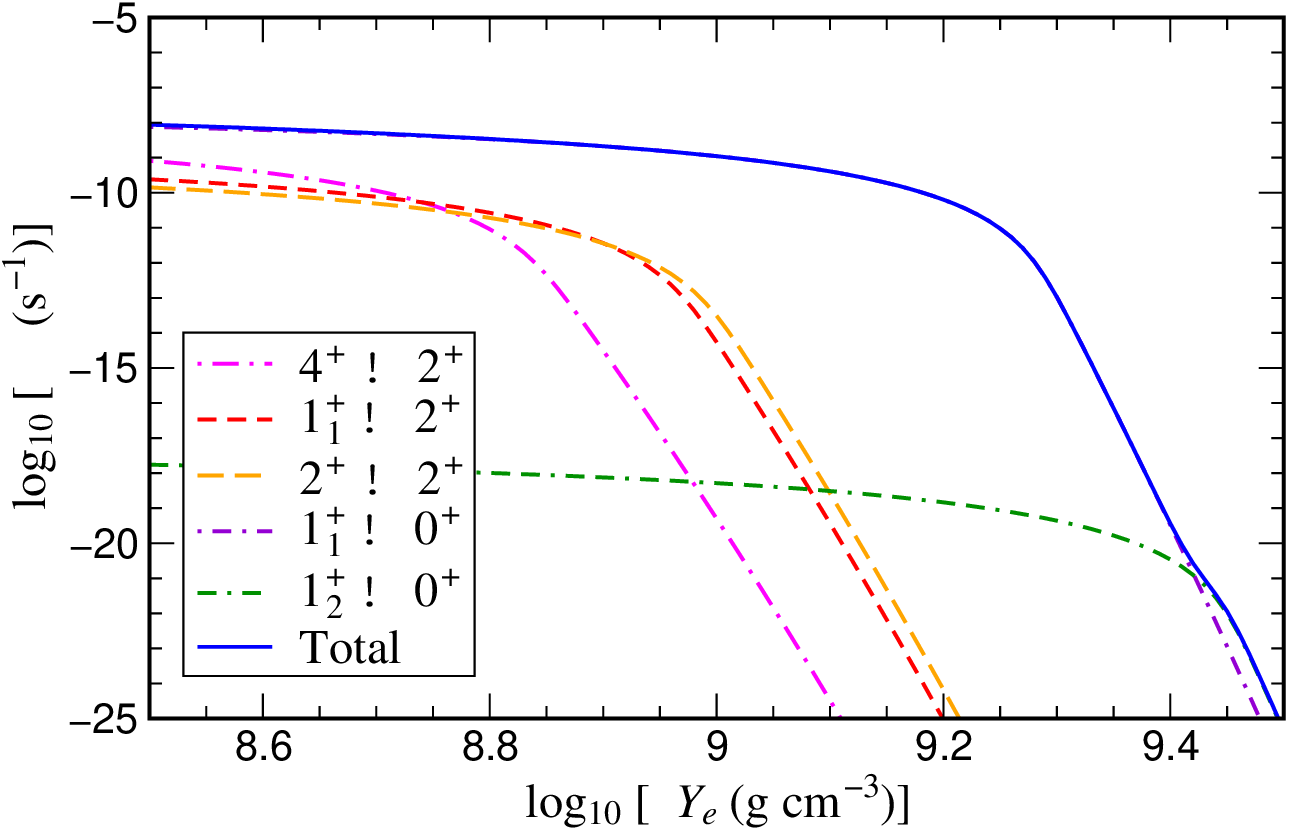}\\
  \includegraphics[width=\linewidth]{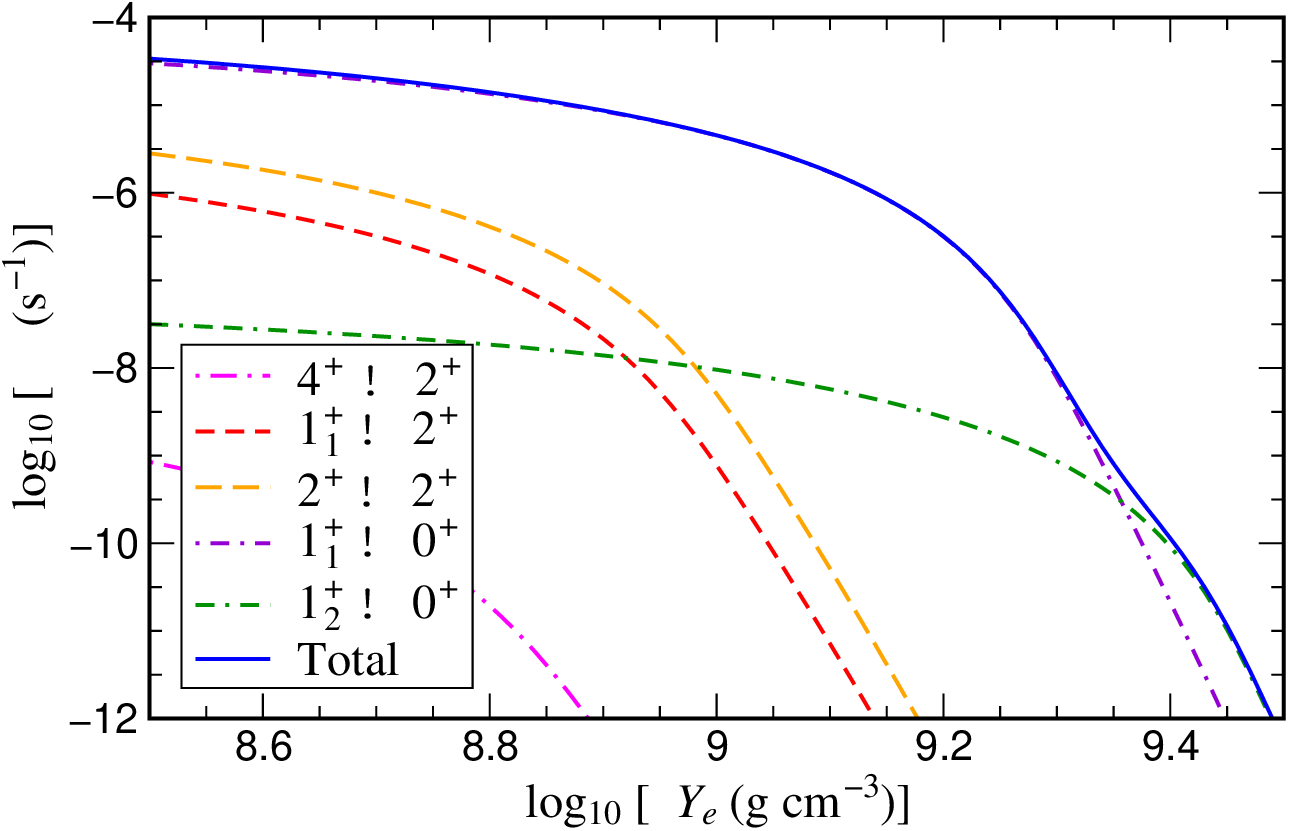}
  \caption{Beta decay of $^{24}$Na as a function of density and for
    selected temperatures (upper panel: $\log\ T(\mathrm{K}) =
    8.6$, lower panel: $\log\ T(\mathrm{K}) = 9.0$).  The figure
    shows the five transitions that fully determine the rate. The
    rates have not been corrected for medium
    effects.\label{fig:na24beta}}
\end{figure}

The present rate is somewhat larger than the one given by Takahara et
al.  \cite{Takahara.Hino.ea:1989} and by Oda et
al. \cite{Oda.Hino.ea:1994}. The main difference comes from the
increase in the B(GT) value for the ground state transition to the
$1^+$ state at 472~keV where recent experiments
\cite{Nishimura.Fujita.ea:2011,Zegers.Meharchand.ea:2008} indicate a
noticeably larger value than deduced from $^{24}$Al decay data
available at the time of the Takahara \emph{et
al.}~\cite{Takahara.Hino.ea:1989} and Oda \emph{et
al.}~\cite{Oda.Hino.ea:1994} works.

Electron capture on $^{24}$Mg stands in competition with the $\beta$
decay of the daughter $^{24}$Na. We have evaluated this $\beta$-decay
rate on the basis of the same transitions as adopted in our
calculation of the $^{24}$Mg electron capture and summarized in
Table~\ref{tab:exp24}.  For the conditions of interest here we find
that the $^{24}$Na $\beta$ decay is mainly given by the GT transition
from the isomeric $1^+$ state at excitation energy $E_x=472$~keV to
the $^{24}$Mg ground state, where the GT strength is known from the
decay of the mirror state in $^{24}$Al. However, this contribution
depends strongly on temperature via the thermal population of the
initial state.  Hence with decreasing temperature the second-forbidden
transition from the $^{24}$Na $4^+$ ground state to the $2^+$ state in
$^{24}$Mg at $E_x=1.369$~MeV grows in importance. This transition
strength is known experimentally from the $\beta$ decay of
$^{24}$Na~\cite{Firestone:2007}. We note that at low temperatures (the
upper panel of Fig.~\ref{fig:na24beta} shows the $\beta$ decay rate at
$T=0.4$ GK) and low densities ($\log \rho Y_e (\textrm{g cm}^{-3}) < 8.75$) this forbidden
transition contributes already of order $10\%$ to the total decay
rate, and becomes relatively more important at even smaller
temperatures. The contribution of this transition has been determined
assuming an allowed shape for the phase space. As discussed for
$^{20}$Ne, it can become even larger if the phase space deviates from
the allowed shape. The GT transition from the isomeric $1^+$ state to
the ground state gets Pauli blocked by the presence of the electron
sea at densities of order ($\log (\rho Y_e) \approx 9.3$), explaining
the strong decrease in its partial rate.  Hence at higher densities
the transition from the second $1^+$ excited state at $E_x=1.347$~MeV
to the ground state, having a larger decay energy, contributes more
strongly to the rate, in particular at the higher temperatures.
Finally we mention that the GT decay from the $^{24}$Na ground state
to the excited state at $E_x=4.123$ MeV in $^{24}$Mg, which
overwhelmingly dominates the $^{24}$Na $\beta$ decay under terrestrial
conditions is strongly Pauli-blocked under astrophysical conditions
with densities $\rho Y_e > 10^8$~g~cm$^{-3}$.

\begin{figure}
  \begin{center}
    \includegraphics[width=\linewidth]{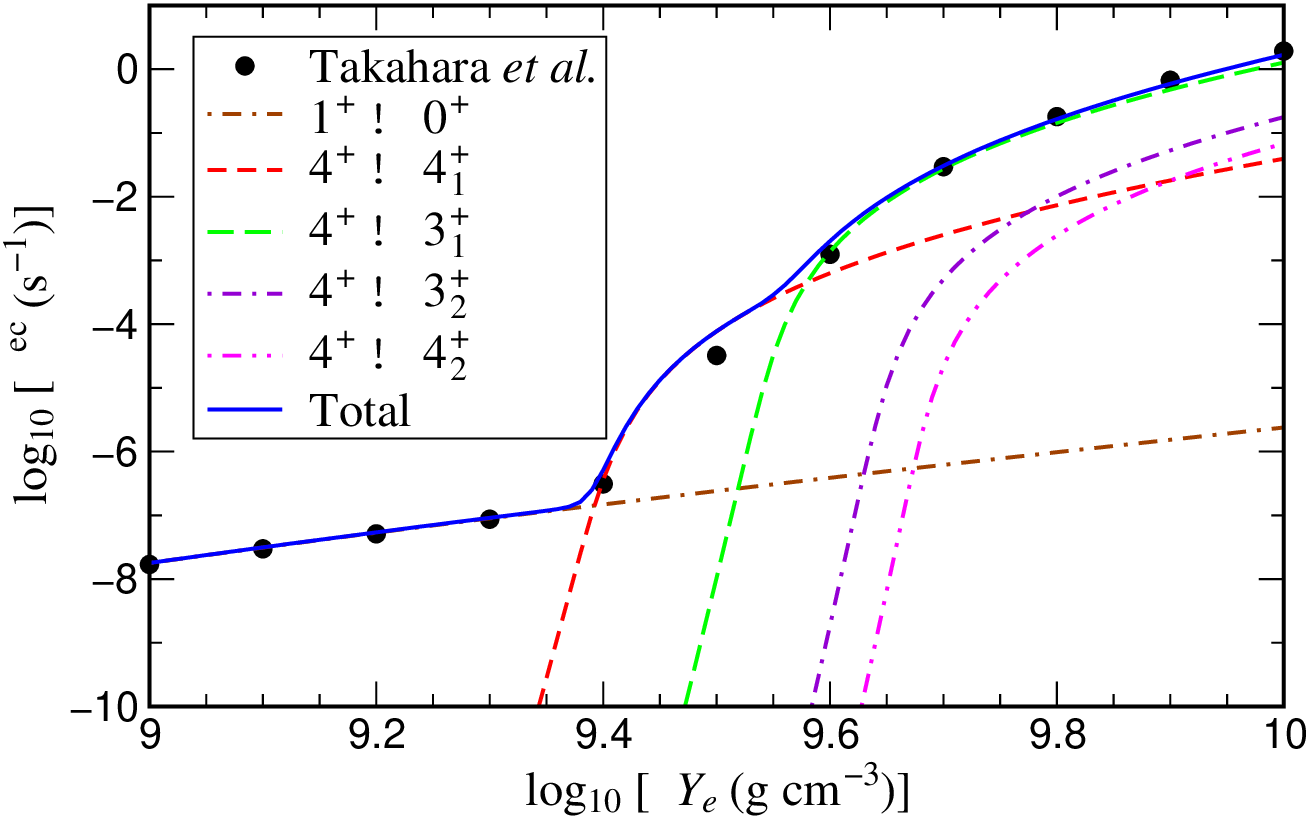}\\
    \includegraphics[width=\linewidth]{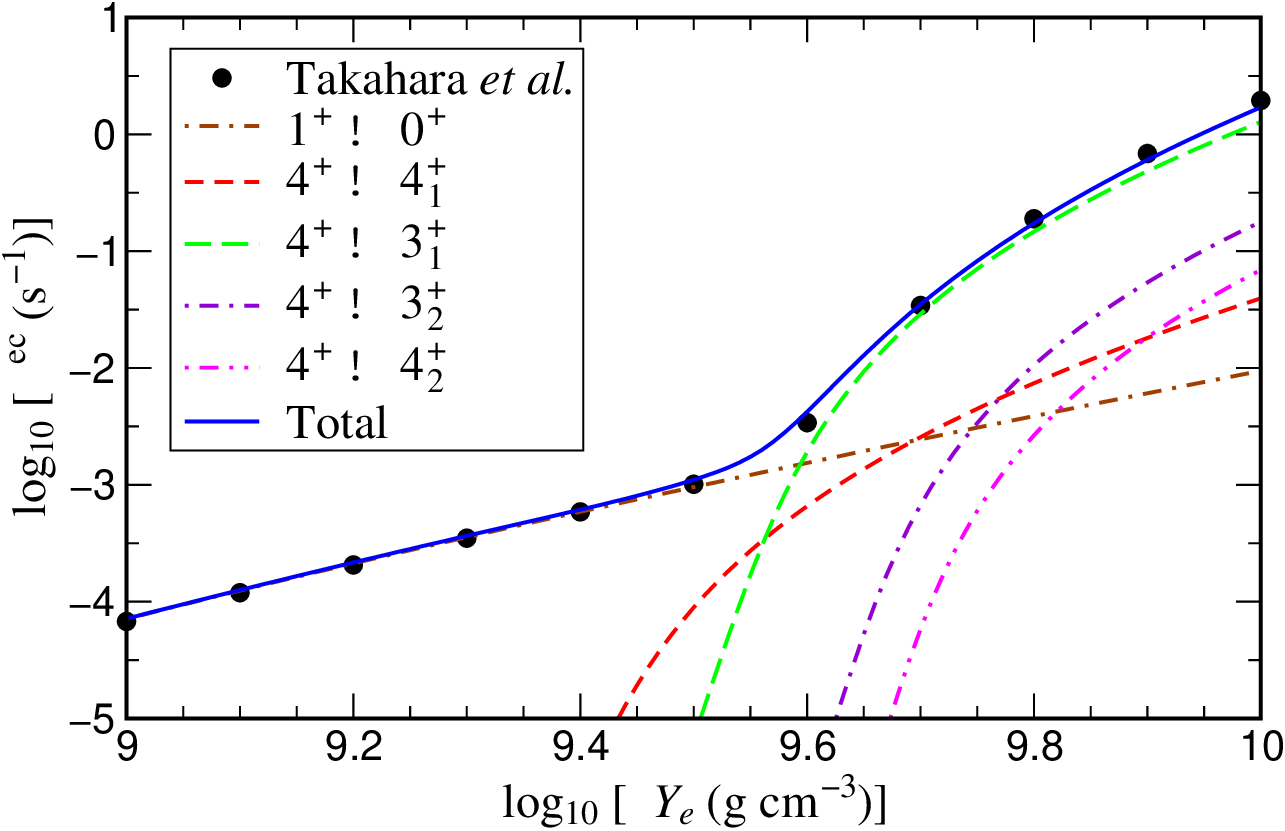}
     \caption{Comparison of our electron capture rate on $^{24}$Na as
      function of density and for selected temperatures (upper panel:
      $\log\ T(\mathrm{K}) = 8.6$, lower panel: $\log\ T(\mathrm{K}) =
      9.0$) with the values given by Takahara \emph{et
        al.}~\cite{Takahara.Hino.ea:1989}. The figure shows the five
      transitions that determine the rate. The rates have not been 
      corrected for medium effects. \label{fig:na24ec}}
  \end{center}
\end{figure}

\begin{table}[hbt]
  \centering
  \caption{Information that determines the 
    electron capture rate on $^{24}$Na. The ground-state to
    ground-state electron $Q$-value is $Q_{\text{ec}} =
    -2.977$~MeV~\cite{Wang.Audi.ea:2012}. \label{tab:expna24}}   
  \begin{ruledtabular}
    \begin{tabular}{ccccc}
      \multicolumn{2}{c}{Initial $^{24}$Na state} &
      \multicolumn{2}{c}{Final $^{24}$Ne state} & Matrix element\\
      \cline{1-2}\cline{3-4} 
      $J^\pi$ & Energy (MeV)& $J^\pi$ & Energy (MeV) & $B$ \\\hline
      $4^+$ & 0 & $4^+_1$ & 3.972 &
      $3.8\times10^{-3}$\footnote{Theoretical value}\\ 
      $4^+$ & 0 & $3^+_1$ & 4.817\footnotemark[1] & 0.232\footnotemark[1]\\
      $4^+$ & 0 & $3^+_2$ & 5.436\footnotemark[1] & 0.058\footnotemark[1]\\
      $4^+$ & 0 & $4^+_2$ & 5.691\footnotemark[1] & 0.030\footnotemark[1]\\
      $1^+$ & 0.472 & $0^+$ &  0.0 & 0.091(2)\footnote{From $^{24}$Ne
        decay~\cite{Alburger:1974}} \\
    \end{tabular}
  \end{ruledtabular}
\end{table}

In Fig.~\ref{fig:na24ec} we have plotted the electron capture rate on
$^{24}$Na at two selected temperatures and the densities of relevance
for the evolution of the ONeMg core in 8--12~M$_\odot$ stars.  We have
evaluated the rate based on experimental energies and shell model GT
transition rates, supplemented by data from the decay of
$^{24}$Ne~\cite{Alburger:1974}. In particular, we include the GT
strengths build on all states till 2.5~MeV excitation energy. The
energies and strengths for the transitions that are relevant to
determine the rate are summarized in Table~\ref{tab:expna24}. We
confirm the results discussed in Ref. \cite{Takahara.Hino.ea:1989}. At
low densities ($\log (\rho Y_e) < 9.4$) the rate is dominated by the
GT transition from the thermally populated isomeric $1^+$ state at
$E_x=472$~keV to the $^{24}$Ne ground state, while at the largest
densities of interest ($\log (\rho Y_e) \sim 9.6$--10) the transition
from the $^{24}$Na $4^+$ ground state to the first excited $3^+$ has
the largest contribution to the total capture rate, with minor
corrections from the GT transition from the ground state to the second
excited $3^+$ state. Our shell model calculations gives quite similar
transition strengths than the one performed by
\cite{Takahara.Hino.ea:1989}. Hence our rates agree quite well with
the previous one in the respective density ranges. However, at
intermediate densities ($\log (\rho Y_e) \sim 9.4$--9.6) the rate is
dominated by the transition from the $^{24}$Na ground state to the
excited $4^+$ state in $^{24}$Ne. For this transition our calculation
predicts a slightly larger GT value than used in
ref.~\cite{Takahara.Hino.ea:1989}, $B=1.5\times 10^{-3}$, explaining
the rate differences in this intermediate regime. (Both, our
calculations based on the USDB~\cite{Brown.Richter:2006} and the
calculations of Takahara \emph{et al.}~\cite{Takahara.Hino.ea:1989}
based on the USD interaction~\cite{Brown.Wildenthal:1988} predict a
very small matrix element and consequently are rather sensitive to the
relatively small differences between the USD and USDB interactions.)
As the relevance of this contribution decreases with increasing
temperature, our $^{24}$Na electron capture rate becomes similar to
the one of Ref.  \cite{Takahara.Hino.ea:1989} at higher temperatures.

As the electron capture and $\beta$-decay rates for the $A=24$ nuclei,
presented here, are dominated by a few states we can derive analytical 
rate expressions, following the procedure as outlined above for the
$A=20$ nuclei. For electron capture on $^{24}$Mg we can write the
electron capture rate, based on the 6 transitions identified above,
as:

\begin{widetext}
  \begin{equation}
  \label{eq:mg24ecana}
  \begin{split}
  \lambda^{\text{ec}}(^{24}\mathrm{Mg}) = \frac{\ln 2}{K}
  \bigl[&5 e^{-E(2^+)/kT} B_e(2^+\rightarrow 1^+) \Phi^{\text{ec}}_e (Q(2^+\rightarrow
    1^+),T,\mu_e) + 5 e^{-E(2^+)/kT} B_e(2^+\rightarrow 2^+)
    \Phi^{\text{ec}}_e (Q(2^+\rightarrow 2^+),T,\mu_e) \\
    & + B_e(0^+\rightarrow1^+_1)\Phi^{\text{ec}}_e(Q(0^+\rightarrow
    1^+_1),T,\mu_e)+ B_e(0^+\rightarrow1^+_2)\Phi^{\text{ec}}_e(Q(0^+\rightarrow
    1^+_2),T,\mu_e)\\
    & + B_e(0^+\rightarrow1^+_3)\Phi^{\text{ec}}_e(Q(0^+\rightarrow
    1^+_3),T,\mu_e)+B_e(0^+\rightarrow1^+_4)\Phi^{\text{ec}}_e(Q(0^+\rightarrow
    1^+_4),T,\mu_e) \bigr].
  \end{split}
\end{equation}
\end{widetext}

\begin{table}[hbt]
  \centering
  \caption{Numerical values of the quantities that determine the
    analytical expression for the electron capture rate on
    $^{24}$Mg. \label{tab:anaMg24ec}}
  \begin{ruledtabular}
    \begin{tabular}{ccc}
      Transition & $Q$-value & Effective matrix element \\
      $J^\pi(^{24}\mathrm{Mg})\rightarrow J^\pi(^{24}\mathrm{Na})$  &
      (MeV)& $B_e$ \\\hline
      $2^+\rightarrow 1^+$ & $-5.129$ & 0.00613 \\
      $2^+\rightarrow 2^+$ & $-5.220$ & 0.0426 \\
      $0^+\rightarrow 1^+_1$ & $-6.498$ & 0.125\\
      $0^+\rightarrow 1^+_2$ & $-7.373$ & 1.383 \\
      $0^+\rightarrow 1^+_3$ & $-7.916$ & 0.0533 \\
      $0^+\rightarrow 1^+_4$ & $-9.436$ & 0.613 \\
    \end{tabular}
  \end{ruledtabular}
\end{table}

The quantities necessary for the evaluation of the rate are defined in
table~\ref{tab:anaMg24ec}. The difference between the effective
transition matrix element, $B_e$, appearing in
table~\ref{tab:anaMg24ec} and the matrix element, $B$, of
table~\ref{tab:exp24} is due to the fact that the former includes the
average value of the Fermi Coulomb distortion function. This value is
equal to 1.332 for $Z=12$.
Equation~(\ref{eq:mg24ecana}) reproduces the electron capture rate on
$^{24}$Mg with a maximum error of about 20\% using the approximate
expressions of the Fermi function defined in
eq.~\eqref{eq:fermiapprox}. The maximum error occurs at densities
$\log(\rho Y_e)=9.3$ where the electron chemical becomes  of the
order of the electron capture $Q$-value, for other conditions the
error is of a few percent.

The beta-decay rate of $^{24}$Na is given by the 5 transitions shown
in figure~\ref{fig:na24beta} for the astrophysical conditions of
interest. The rate can then be written as:

\begin{widetext}
\begin{equation}
  \label{eq:na24beana}
  \begin{split}
  \lambda^{\beta}(^{24}\mathrm{Na}) = \frac{\ln 2}{9 K}
  \bigl[&9 B_e(4^+\rightarrow 2^+) \Phi^{\beta}_e (Q(4^+\rightarrow 
    2^+),T,\mu_e) + 3 e^{-E(1^+_1)/kT} B_e(1^+_1\rightarrow 2^+)
    \Phi^{\beta}_e (Q(1^+_1\rightarrow  2^+),T,\mu_e)\\
    &+5 e^{-E(2^+)/kT} B_e(2^+\rightarrow 2^+) \Phi^{\beta}_e (Q(2^+\rightarrow 
    2^+),T,\mu_e)
    +3 e^{-E(1^+_1)/kT} B_e(1^+_1\rightarrow 0^+)\Phi^{\beta}_e(Q(1^+_1\rightarrow 
    0^+),T,\mu_e)\\
    &+3 e^{-E(1^+_2)/kT} B_e(1^+_2\rightarrow 0^+)\Phi^{\beta}_e(Q(1^+_2\rightarrow 
    0^+),T,\mu_e)\bigr].
  \end{split}
\end{equation}
\end{widetext}

\begin{table}[hbt]
  \centering
  \caption{Numerical values of the quantities that determine the
    analytical expression for the beta decay rate of
    $^{24}$Na. \label{tab:anaNa24bet}}
  \begin{ruledtabular}
    \begin{tabular}{ccc}
      Transition & $Q$-value & Effective matrix element \\
      $J^\pi(^{24}\mathrm{Mg})\rightarrow J^\pi(^{24}\mathrm{Na})$  &
      (MeV)& $B_e$ \\\hline
      $4^+\rightarrow 2^+$ & $4.657$ & $3.70\times 10^{-8}$ \\
      $1^+_1\rightarrow 2^+$ & $5.129$ & 0.0102 \\
      $2^+\rightarrow 2^+$ & $5.220$ & 0.0426 \\
      $1^+_1\rightarrow 0^+$ & $6.498$ & 0.0417\\
      $1^+_2\rightarrow 0^+$ & $7.373$ & 0.461 \\
    \end{tabular}
  \end{ruledtabular}
\end{table}

The quantities necessary for the evaluation of the rate are defined in
table~\ref{tab:anaNa24bet}. Equation~(\ref{eq:na24beana}) reproduces
the beta-decay rate of $^{24}$Na with a maximum error of around 20\%
using the approximate expressions of the Fermi function defined in
eq.~\eqref{eq:fermiapprox}. The maximum error occurs at densities
$\log(\rho Y_e)=9.3$ where the electron chemical becomes of the order
of the electron capture $Q$-value, for other conditions the error is
of a few percent.

Finally, the electron capture rate on $^{24}$Na can be written as: 
\begin{widetext}
\begin{equation}
  \label{eq:na24ecana}
  \begin{split}
  \lambda^{\text{ec}}(^{24}\mathrm{Na}) = \frac{\ln 2}{9 K}
  \bigl[&3 e^{-E(1^+)/kT} B_e(1^+\rightarrow 0^+) \Phi^{\text{ec}}_e
  (Q(1^+\rightarrow  
    0^+),T,\mu_e) + 9 B_e(4^+\rightarrow 4^+_1)
    \Phi^{\text{ec}}_e (Q(4^+\rightarrow 4^+_1),T,\mu_e)\\ 
    &+ 9 B_e(4^+\rightarrow 3^+_1) \Phi^{\text{ec}}_e (Q(4^+\rightarrow
    3^+_1),T,\mu_e) + 9 B_e(4^+\rightarrow 3^+_2) \Phi^{\text{ec}}_e
    (Q(4^+\rightarrow  3^+_2),T,\mu_e)\\
    &+ 9 B_e(4^+\rightarrow 4^+_2) \Phi^{\text{ec}}_e (Q(4^+\rightarrow
    4^+_2),T,\mu_e) \bigr]
  \end{split}
\end{equation}
\end{widetext}

\begin{table}[hbt]
  \centering
  \caption{Numerical values of the quantities that determine the
    analytical expression for the electron capture rate on
    $^{24}$Na. \label{tab:anaNa24ec}}
  \begin{ruledtabular}
    \begin{tabular}{ccc}
      Transition & $Q$-value & Effective matrix element \\
      $J^\pi(^{24}\mathrm{Na})\rightarrow J^\pi(^{24}\mathrm{Ne})$  &
      (MeV)& $B_e$ \\\hline
      $1^+\rightarrow 0^+$ & $-2.505$ & 0.118 \\
      $4^+\rightarrow 4^+_1$ & $-6.949$ & $4.94\times 10^{-3}$ \\
      $4^+\rightarrow 3^+_1$ & $-7.794$ & 0.301 \\
      $4^+\rightarrow 3^+_2$ & $-8.413$ & 0.0753\\
      $4^+\rightarrow 4^+_2$ & $-8.668$ & 0.0390 \\
    \end{tabular}
  \end{ruledtabular}
\end{table}

The quantities necessary for the evaluation of the rate are defined in
table~\ref{tab:anaNa24ec}. The difference between the effective
transition matrix element, $B_e$, appearing in
table~\ref{tab:anaNa24ec} and the matrix element, $B$, of
table~\ref{tab:expna24} is due to the fact that the former includes
the average value of the Fermi Coulomb distortion function. This value
is equal to 1.299 for $Z=11$. Equation~(\ref{eq:na24ecana}) reproduces
the beta-decay rate of $^{24}$Na with a maximum error of around 10\%
at temperatures around $\log\ T(\mathrm{K}) = 8.6$ using the
approximate expressions of the Fermi function defined in
eq.~\eqref{eq:fermiapprox}. The maximum error occurs at densities
$\log(\rho Y_e)=9.4$ where the electron chemical becomes of the order
of the electron capture $Q$-value for the transition $4^+\rightarrow
4^+_1$. With increasing temperature or for other densities the error
is of a few percent.

\begin{figure}[htb]
  \begin{center}
    \includegraphics[width=\linewidth]{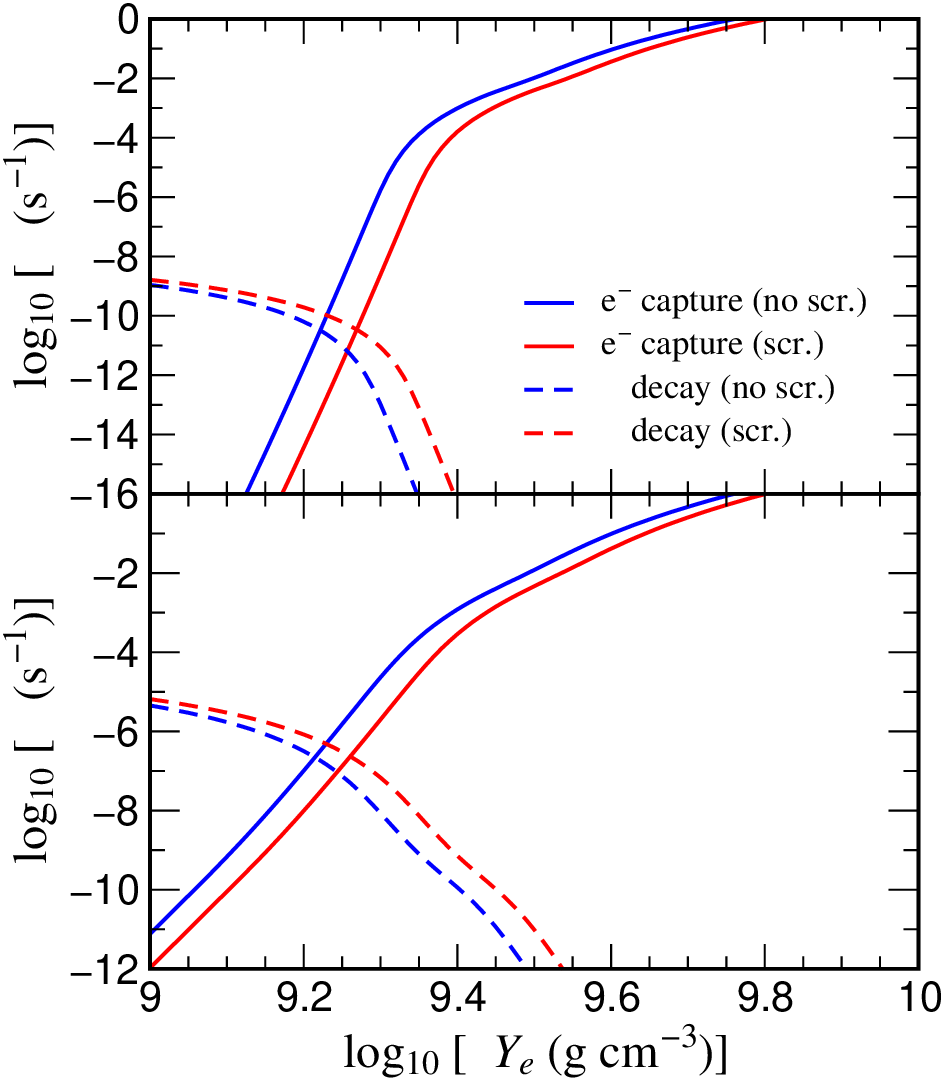}
    \caption{Electron capture rate on $^{24}$Mg and beta decay rate of
      $^{24}$Na with and without consideration of medium corrections
      (upper panel: $\log\ T(\mathrm{K}) = 8.6$, lower panel: $\log\
      T(\mathrm{K}) = 9.0$).\label{fig:mg24na-scr}}
  \end{center}
\end{figure}

As explained above medium corrections decrease the electron capture
rates and increase the competing $\beta$-decay rates. This is visible
in Fig.~\ref{fig:mg24na-scr} where we compare the electron capture
rate on $^{24}$Mg and the $\beta$-decay rate of $^{24}$Na calculated
with and without the Coulomb corrections, applying the same formalism
as discussed above for the case of the $A=20$ nuclei. As for the
$^{20}$Ne-$^{20}$F pair, the Coulomb corrections also shift the
density at which the $^{24}$Mg capture and the $^{24}$Na decay rate
become identical towards higher values, again by about 0.05 dex in
$\log(\rho Y_e)$. We expect that this shift is a typical value for
sd-shell nuclei at the conditions of the collapsing ONeMg core and
should hence also affect the URCA pairs ($^{23}$Na-$^{23}$Ne,
$^{25}$Mg-$^{25}$Na, $^{27}$Al-$^{27}$Mg) which play an important role
for the cooling during the collapse~\cite{Toki.Suzuki.ea:2013}.

\begin{figure}[htb]
  \centering
  \includegraphics[width=\linewidth]{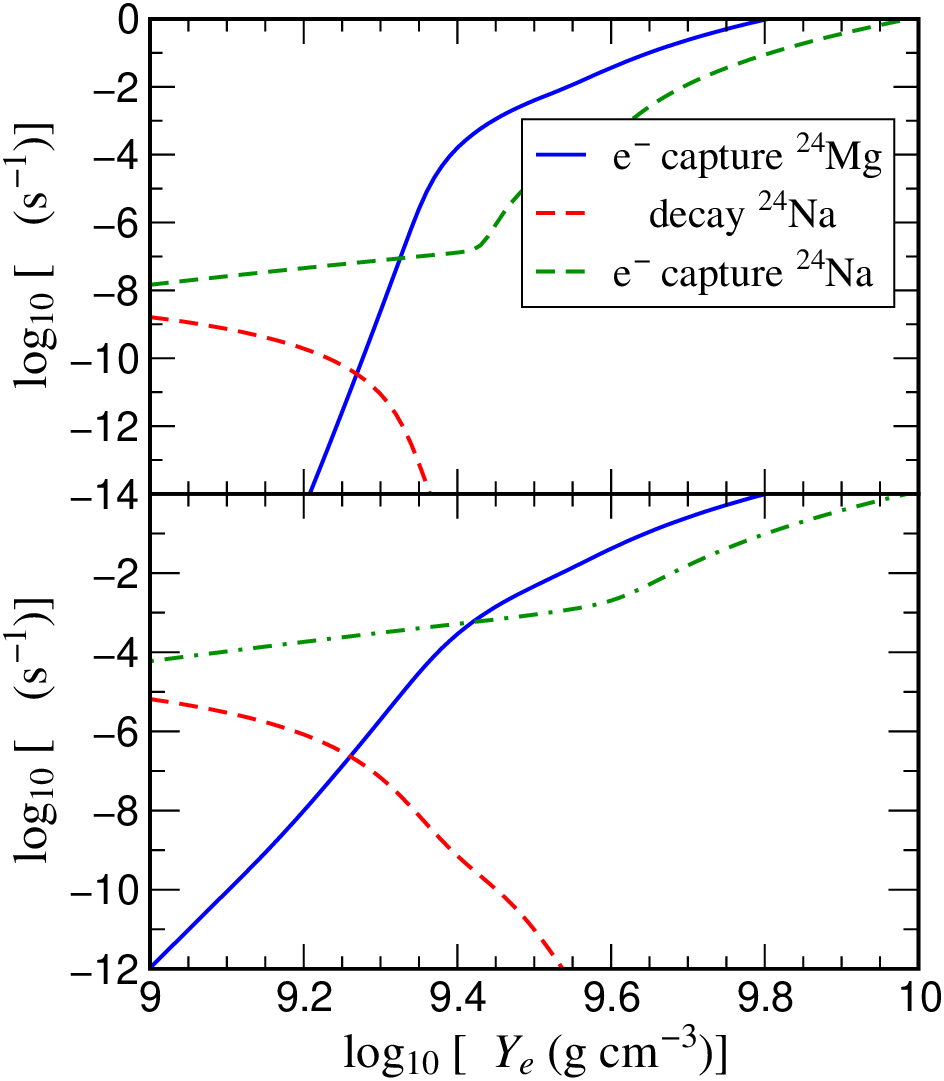}
  \caption{Rates for electron captures on $^{24}$Mg and $^{24}$Na and
    beta-decay of $^{24}$Na for selected temperatures: $\log\
    T(\mathrm{K}) = 8.6$ (upper panel), $\log\ T(\mathrm{K}) = 9.0$
    (lower panel). Medium corrections are included in the
    rates. \label{fig:mg24na24}} 
\end{figure}

In Fig. \ref{fig:mg24na24} we compare the medium-corrected rates for
the $A=24$ nuclei and observe that electron capture on $^{24}$Mg
dominates over $^{24}$Na $\beta$ decay for densities $\log \rho Y_e >
9.3$.  However, at the astrophysical conditions at which the electron
capture rate on $^{24}$Mg is larger than the competing $^{24}$Na
$\beta$ decay, it is also larger than the capture rate on the daughter
$^{24}$Na. The consequence is that once capture on $^{24}$Mg is faster
than $\beta$ decay it is followed by a second capture process leading
to $^{24}$Ne.

\begin{figure}[htb]
  \centering
  \includegraphics[width=\linewidth]{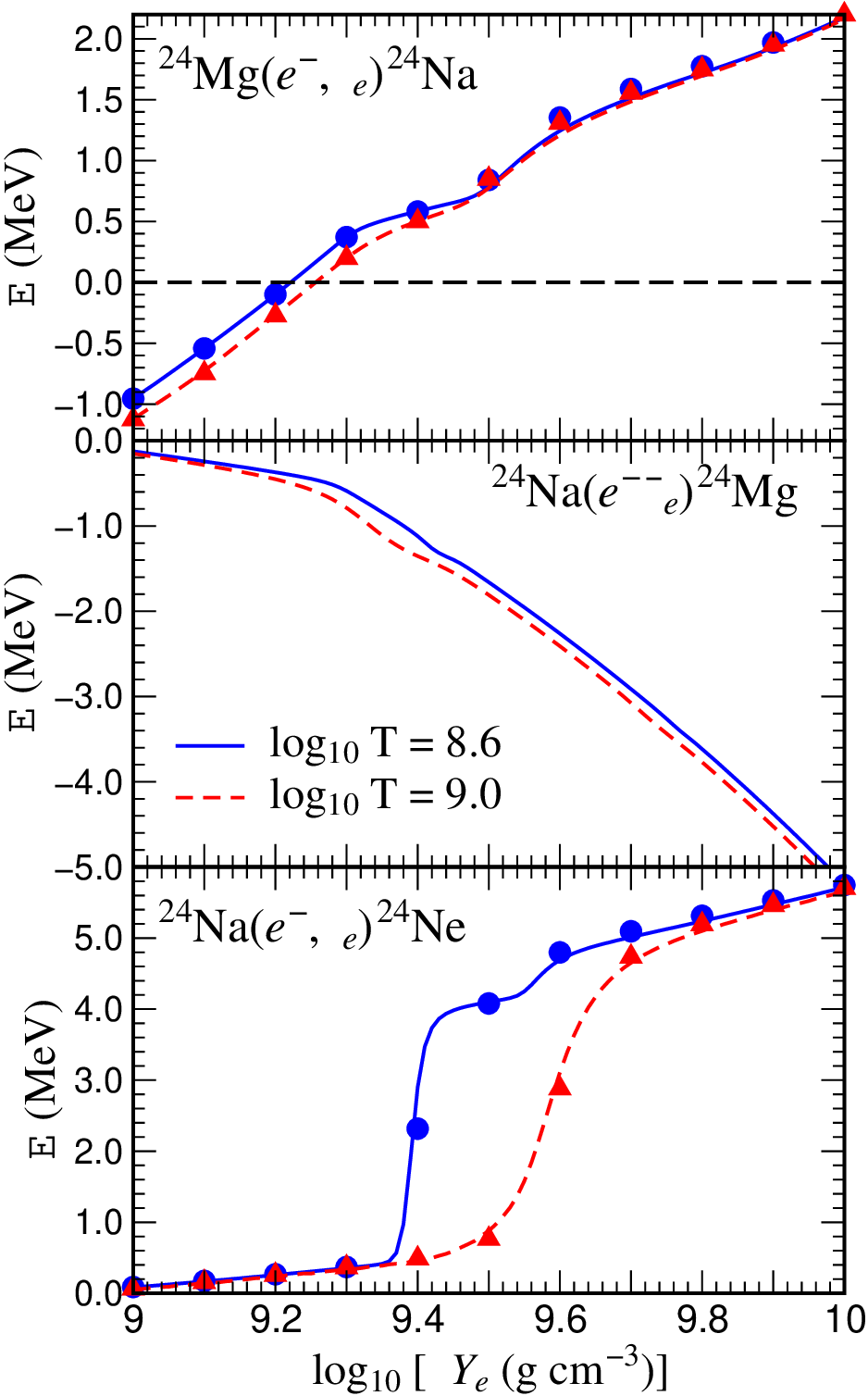}
  \caption{Average energy produced (positive) or absorbed (negative)
    by electron capture on $^{24}$Mg (upper panel), beta-decay of
    $^{24}$Na (middle panel) and electron capture on $^{24}$Na (lower
    panel).\label{fig:energe24}}
\end{figure}

In Fig.~\ref{fig:energe24} we plot the quantity $\mathcal{E}$ for
electron capture on $^{24}$Mg and $^{24}$Na and for $^{24}$Na $\beta$
decay. The capture on $^{24}$Mg is endothermic up to about $\log (\rho
Y_e) = 9.3$, while the capture process is exothermic at densities
$\rho Y_e > 10^9$~g~cm$^{-3}$. Like for the $A=20$ nuclei this
difference is due to pairing energy making the $Q$ value for capture
on $^{24}$Mg larger than for capture on $^{24}$Na. The $^{24}$Na
$\beta$ decay is endothermic, explaining the negative values of
$\mathcal{E}$.  The kinks in $\mathcal{E}$ occur when the rate is
changing from the dominance of one specific transition to another.
The respective transitions are identified in our discussion of the
individual rates. We observe good agreement with the results obtained
by Takahara \emph{et al.} \cite{Takahara.Hino.ea:1989} for the
electron capture processes (these authors do not give results for the
$\beta$ decay).  For densities $\log(\rho Y_e)\ (\textrm{g~cm}^{-3})
>9.2$, the energy produced by capture on $^{24}$Mg becomes
positive. This together with the fact that capture on $^{24}$Na always
produce energy for the conditions considered marks the transition
density at which the net effect of the sequence of $A=24$
weak-interaction processes changes from endothermic to exothermic
increasing the core temperature in stellar evolution models.

\section{Conclusion}

We have calculated the rates for electron captures on $^{20}$Ne,
$^{20}$F, $^{24}$Mg, $^{24}$Na and $\beta$ decays of $^{20}$F and
$^{24}$Na which are key quantities for studies of the late-time
evolution of 8--12~M$_\odot$ stars.  So far such late-time studies are
based on the rate evaluations of~\cite{Takahara.Hino.ea:1989}
and~\cite{Oda.Hino.ea:1994}. We have improved these rates in three
important aspects. First, we have incorportated experimental data from
either $\beta$ decay or charge-exchange experiments which have not
been available at the time when Takahara \emph{et
  al.}~\cite{Takahara.Hino.ea:1989} and Oda \emph{et
  al.}~\cite{Oda.Hino.ea:1994} did their work. In our study the recent
experimental data are supplemented by Gamow-Teller transitions derived
from large-scale shell model calculations, similar to the procedure
in~\cite{Takahara.Hino.ea:1989,Oda.Hino.ea:1994}. Importantly we find 
that nuclear physics input into the astrophysically relevant rates for
electron captures on $^{20}$Ne and $^{24}$Mg and the competing beta
decays of the respective daughters is completely based on experimental
data. The exception is the electron capture on $^{20}$Ne in the
density regime $\log \rho Y_e = 9.3$--9.7. As our second improvement
we point out that at temperatures $T < 0.7 \times 10^9$ the capture
rate is likely to be dominated by the second-forbidden transition from
the $^{20}$Ne ground state to the $^{20}$F. Experimentally only an
upper limit exists, which, however, is of the order of typical
second-forbidden transition strengths. While we have used the upper
limit as an estimate for this transition in our present work a
calculation of the transition with an appropriate method like the
shell model or an experimental determination is highly desirable.

As the third improvement, we have corrected the various rates for
medium-induced effects. Here we followed the formalism discussed in
ref.~\cite{Juodagalvis.Langanke.ea:2010} for electron captures and
extended it to the treatment for $\beta$-decays. The environment
reduces the electron chemical potential and enhances (reduces) the
reaction $Q$-value for electron captures ($\beta$ decays). As a
consequence electron capture rates are lower in dense astrophysical
environments than for bare nuclei, while $\beta$ decay rates are
larger.  For the astrophysical conditions here, the medium corrections
change the rates typically by a factor of order two. The effect is, of
course, significantly larger at such densities where the rates change
from dominance of a certain transition to another (as is the case in
the weak processes here) as these transitions are extremely sensitive
to the effective $Q$ values.

We note that medium effects should also have a significant effect on
the densities at which so-called URCA pairs operate and influence the
late-stage evolution of the stars. As $\beta$ decay rates are enhanced
and the competing electron capture rates are lowered, the medium
modifiactions will move the operation of the URCA pairs to somewhat
higher densities. As the shifts of the electron chemical potential and
of the $Q$ values are of order 100~keV under the relevant density (and
temperature) conditions encountered, we expect that the URCA pairs
operate at densities which are about $0.1 \times 10^9$~g~cm$^{-3}$ larger
than found in calculations which do not consider medium corrections on
the rate. Stellar evolution studies which investigate the impact of
screening on the URCA pairs are needed.

We have presented analytical expressions for both electron capture and
beta-decay rates that allow for an accurate description of these
processes for conditions at which URCA pairs operate in both
intermediate mass stars~\cite{Tsuruta.Cameron:1970} and neutron
stars~\cite{Schatz.Gupta.ea:2014}. 

Rate tables on fine grids in temperature and density in the ranges
$\rho Y_e = 10^8 - 10^{10}$~g~cm$^{-3}$ and $T=10^8 - 10^{10}$~K can be
obtained by request from the authors.

\section*{Acknowledgements}
\label{sec:acknowledgements}

This work was supported by the ExtreMe Matter Institute EMMI in the
framework of the Helmholtz Alliance HA216/EMMI, the Deutsche
Forschungsgemeinschaft through contract SFB~634, the Helmholtz
International Center for FAIR within the framework of the LOEWE
program launched by the state of Hesse, the Helmholtz Association
through the Nuclear Astrophysics Virtual Institute (VH-VI-417) and
USNSF (PHY-1102511, PHY-0822648(JINA)). We thank B. A. Brown,
T. Fischer, S. Jones, R. Hirschi, P. von Neumann-Cosel, and H. Schatz
for fruitful discussions.

\bibliography{../biblio/bibliography}

\end{document}